\documentclass[twocolumn]{emulateapj}
\usepackage{graphicx}
\usepackage{natbib}
\usepackage{float}
\restylefloat{table}
\usepackage{amsmath}
\usepackage{mathtools}
\usepackage{hyperref}
\DeclarePairedDelimiter\abs{\lvert}{\rvert}%

\usepackage[ruled, noend, slide]{algorithm2e}

\newcommand{\quotes}[1]{``#1''}

\begin{document}

\renewcommand{\arraystretch}{1.2}

\title{Clustering based feature learning on variable stars.}

 \author{Crist\'obal Mackenzie \altaffilmark{1}, Karim Pichara \altaffilmark{1}\altaffilmark{,2}\altaffilmark{,4}, 
 Pavlos Protopapas \altaffilmark{2}\altaffilmark{,3}}
\affil{\altaffilmark{1}Computer Science Department, Pontificia Universidad Cat\'olica de Chile, Santiago, Chile}
\affil{\altaffilmark{2}Institute for Applied Computational Science, Harvard University, Cambridge, MA, USA}
\affil{\altaffilmark{3}Harvard-Smithsonian Center for Astrophysics, Cambridge, MA, USA}
\affil{\altaffilmark{4}Millennium Institute of Astrophysics, Chile}
%\affil{\altaffilmark{4}The Milky Way Millennium Nucleus, Av. Vicu\~{n}a Mackenna 4860, 782-0436 Macul, Santiago, Chile}

%\pagerange{\pageref{firstpage}--\pageref{lastpage}} \pubyear{2012}
%\label{firstpage}

\begin{abstract}
The success of automatic classification of variable stars strongly depends on the lightcurve representation. Usually, lightcurves are represented as a vector of many statistical descriptors designed by astronomers called features. These descriptors commonly demand significant computational power to calculate, require substantial research effort to develop and do not guarantee good performance on the final classification task. Today, lightcurve representation is not entirely automatic; algorithms that extract lightcurve features are designed by humans and must be manually tuned up for every survey. The vast amounts of data that will be generated in future surveys like LSST mean astronomers must develop analysis pipelines that are both scalable and automated. Recently, substantial efforts have been made in the machine learning community to develop methods that prescind from expert-designed and manually tuned features for features that are automatically learned from data. In this work we present what is, to our knowledge, the first unsupervised feature learning algorithm designed for variable objects. Our method works by extracting a large number of lightcurve subsequences from a given set of photometric data, which are then clustered to find common local patterns in the time series. Representatives of these common patterns, called exemplars, are then used to transform lightcurves of a labeled set into a new representation that can then be used to train an automatic classifier. The proposed algorithm learns the features from both labeled and unlabeled lightcurves, overcoming the bias generated when the learning process is done only with labeled lightcurves. We test our method on MACHO and OGLE datasets; the results show that the classification performance we achieve is as good and in some cases better than the performance achieved using traditional statistical features, while the computational cost is significantly lower. With these promising results, we believe that our method constitutes a significant step towards the automatization of the lightcurve classification pipeline.
\end{abstract}


\newpage

\section{Introduction}
Automatic classification of variable stars has received substantial attention in the research community in the last years \citep{Debosscher:2007, Wachman:2009, kim2009detrending, Wang:2010, Richards:2011, bloom2011data, Kim:2011, Pichara_QSO:2012, bloom2012automating, Pichara_Miss:2013, kim2014epoch, nun2014supervised, masci2014automated, hanif2015recursive, neff2015automated, babu2015skysurveys}. Achieving a good performance with these methods depends strongly on the way lightcurves are represented. Lightcurves are commonly represented as a vector of many statistical descriptors called features, which aim to measure a particular characteristic of the lightcurve. Feature calculation is intensive in computing resources, the development of new features requires a lot of research effort and new features do not guarantee better classification performance. The upcoming and ongoing deep-sky surveys such as Pan-STARRS \citep{Kaiser2002SPIE}, LSST \citep{Matter:2007} and SkyMapper \citep{Keller:2007} are creating immense amounts of data, which makes automatic and scalable analysis tools an important task for the astronomical community. Today, lightcurve representation is not entirely automatic: algorithms that extract lightcurve features are designed by astronomers and have to be manually tuned up every time new surveys are coming. 

Most of the automatic classification tools coming from the Machine Learning community are very effective in the sense that they can produce high accuracy results and work very fast in the classification stage (after the training phase). However, classification algorithms results are highly dependent on the way the data is represented, and a lot of effort is put in designing features to express lightcurves. For example, \citet{Kim:2011} used a Support Vector Machine (SVM) to classify variable stars, previously defining a set of time series descriptors to be used as features in the classification model. In a later work, \citet{Pichara_QSO:2012} made an important improvement in accuracy thanks to the inclusion of new features coming from a Continuous Auto Regressive model. \citet{huijse2012information} made an improvement in periodic star classification by using an information theoric approach to estimate periodicities. \citet{nun2014supervised} devised a method to detect anomalies in astronomical catalogs by using the results of a Random Forest classification as input for a Bayesian Network.
 
The data representation problem arises in most fields that deal with data like time series and images since the complexity and size of the data usually make it unsuitable as direct input to any classification algorithm. To deal with this issue, the machine learning community propose a new way of representing data: unsupervised feature learning. This method aims to use unlabeled data to learn a model that can be then used to transform data of the same kind to a new representation suitable for classification tasks. This process of transforming data from its raw form to another is known as encoding. The development of unsupervised feature learning started with the objective of finding a good representation of images that could serve as input for learning algorithms. While the goal in most works is similar, approaches vary in nature. \citet{olshausen1996emergence} use sparse coding to represent an image, \citet{bell1997independent} base their approach on signal analysis, \citet{hinton2006reducing} use models based on neural networks, while \citet{coates2012learning} follow a clustering-based method.

We build on the ideas in \citet{coates2012learning} even though their proposed method is not well established for time series. We make substantial modifications to their approach to get meaningful results while using lightcurve data instead of images. These modifications have resulted in a new unsupervised learning method for lightcurves and time series in general. Our method is based on the clustering of hundreds of thousands of lightcurve subsequences, which allows us to find the most common and representative patterns in large amounts of data. The results of the clustering step are then used to transform lightcurves of a labeled set to a representation suitable for machine learning algorithms.

The purpose of this work is to introduce unsupervised feature learning as a strong alternative to expert-designed features that have traditionally been used for lightcurve representation in the context of automatic classification. The performance of classification models trained with data from our method is as good and some cases better than classifiers trained using the traditional lightcurve representation. 

The remainder of this paper is organized as follows: Section \ref{sec:related_work} gives an account of the previous work in feature design for variable stars and the field of unsupervised feature learning, Section \ref{sec:background} introduces the relevant background theory for this work, Section \ref{sec:method}  gives a detailed account of our methodology and Section \ref{sec:data} presents the lightcurve catalogs and training sets used in this work. Section \ref{sec:implementation} discusses some implementation details, and we show our results in Section \ref{sec:results}. We give a brief run-time analysis in Section \ref{sec:computational_cost} and state the conclusions of our work in Section \ref{sec:conclusions}.

\section{Related Work}
\label{sec:related_work}

Automatic classification of lightcurves is currently performed by first transforming each lightcurve to a vector of many statistical descriptors, commonly called features, and then by training a learning algorithm. These features try to capture characteristics related to variability and periodicity, amongst others. \citet{Debosscher:2007} represented lighcurves as a vector of 28 parameters derived mainly from periodicity analysis. \citet{kim2009detrending} introduced the Anderson-Darling test in their method to de-trend lightcurves, which tests whether a given lightcurve can be said to be drawn from a Normal distribution. This test has been included as a lightcurve feature in later work. \citet{Richards:2011} introduced features that measure aspects like kurtosis, skewness, amplitude, deviation from the mean magnitude, linear slope and many features extracted from periodicity analysis using the Lomb-Scargle periodogram. \citet{Kim:2011} designed features to measure variability and dispersion and introduced the use of two photometric bands for some calculations. \citet{pichara2012improved} proposed the use of the continuous auto-regressive model to strengthen the analysis of irregularly sampled lightcurves. \citet{huijse2012information} estimated periodicities with an algorithm based on information theory. \citet{kim2014epoch} introduced more features that relate variability and quartile analysis. \citet{nun2015fats} designed a library that aims to facilitate feature extraction for astromonical lightcurves which includes a compendium of features utilised throughout the recent literature. The design of all features for lightcurve representation that exist today has been the result of many years of research effort.

The tremendous amount of effort required to design new features has driven the focus of many research communities tackling other classification problems away from feature design and towards an unsupervised feature learning approach. Unsupervised feature learning models first emerged in the computer vision community as an effort to find a compact vector representation of  images \citep{olshausen1996emergence}. Many of the models have since been adapted to work with time series data like speech, music, stock prices and sensor readings. The results are varied with some unsupervised learning approaches clearly improving the state-of-the-art performance on benchmark datasets. Sparse Coding \citep{olshausen1996emergence, lee2006efficient}, a methodology that aims to learn a set of over-complete basis which can be used to represent data efficiently, was used by \citet{Grosse07} for audio classification. Another common model that has been employed to solve time series problems is the Restricted Boltzmann Machine \citep{hinton2006reducing, hinton2006fast, larochelle2008classification}. The Restricted Boltzmann Machine (RBM) is a model that learns a distribution over its input data and is represented by an undirected bipartite graph. The weight matrix $W$, which describes the connections between nodes in the graph can be used to transform data to lower dimensional representation. This model has been used with success as a replacement for Gaussian mixtures in the discretization step required for Hidden Markov Models for audio classification \citep{mohamed2012acoustic, dahl2012context}. \citet{jaitly2011learning} used raw speech data as input for an RBM with success. Some variations of the RBM like the mean-covariance RBM \citep{ranzato2010modeling, krizhevsky2010factored} also have been used to improve on audio classification benchmarks \citep{dahl2010phone}. 

Other somewhat less popular unsupervised feature learning models that have been used with success in time series problems are the Recurrent Neural Network \citep{husken2003recurrent}, the Autoencoder \citep{poultney2006efficient, hinton2006reducing, bengio2009learning} and clustering approaches \citep{coates2012learning}. The Recurrent Neural Network (RNN) is essentially a neural network in which the outputs are connected back to the inputs. It has been used with success in replacing both the Gaussian mixture and the Hidden Markov Model in the traditional audio classification pipeline \citep{graves2013speech}. The Autoencoder (AE) is a neural network that tries to model the identity function of its input data. The weights in the network are adjusted during training to make the network's output as close as possible to its input. \citet{langkvist2012not} use a modified version of the AE to perform unsupervised feature learning on sensor data, outperforming the best classification results obtained with expert-designed features. In clustering-based unsupervised feature learning, data is transformed into a new representation as a function of both the data and the most common data patterns found during clustering. \citet{nam2012learning} employ a clustering based approach in combination with other models to perform music classification.

Due to the complexity of time series data, most of the works listed above still tackle unsupervised feature learning with the aid of some form of pre-processing which requires both computational time and domain expertise. Raw time series data has been used with success in a limited number of problems, most notably by \citet{jaitly2011learning}. The previously mentioned models, on the other hand, are not designed to deal with the kind of time series that are common in astronomical surveys. Lightcurves are not sampled uniformly, so they have different number of observations for a fixed time frame. These characteristics of the data make it unsuitable as input for neural network based models like the RBM and the AE, sparse coding, and most models that assume that the input is a vector of a fixed size. Sensor data, digital sound, and stock prices do not have this problem, since they are sampled uniformly. Given the massive amounts of astronomical data to be collected in future surveys, the development of an automated pipeline for raw data analysis with minimal pre-processing is a priority.

\section{Background Theory}
\label{sec:background}

Unsupervised feature learning algorithms, like the ones described in the previous section, work by learning a model from the usually vast amounts of unlabeled data available which can then be used to transform data to a representation suitable for machine learning tasks. The way we model the data in our feature learning approach is through a large set of representative local patterns that cover common occurrences in the lightcurves. To find these patterns, we run a clustering algorithm on a large set of unlabeled lightcurve subsequences, and then consider the representatives of each cluster found as a pattern to be included in our model.

When clustering any data, the measure used to evaluate the similarity between data points is of extreme importance to the quality of the results. In the domain of time series, the use of the standard similarity measures like Euclidean distance and $L_P$ norms, in general, is not suitable. Astronomical lightcurves are unevenly sampled and thus, the time series under comparison are rarely of the same length, so the Euclidean distance is not even well defined for the comparison of this kind of data. To solve this problem, \quotes{elastic measures} that tolerate uneven sampling and time series of different length have been proposed \citep{berndt1994using, chen2005robust}. \citet{serra2014empirical} have found the Time Warp Edit Distance \citep{marteau2009time} to be one of the most powerful and flexible for the case of unevenly sampled time series. Given that it allows for a meaningful comparison between any pair of time series of different length with even or uneven sampling, we use the Time Warp Edit Distance as the similarity measure for lightcurves in our experiments. The Time Warp Edit Distance is based on the Levenshtein Distance \citep{levenshtein1966binary}, commonly known as Edit Distance, which was initially defined as a measure to assess the similarity between two strings of characters and has been adapted to work with time series.

The use of the Time Warp Edit Distance as the similarity measure for lightcurve comparison poses an additional challenge for our lightcurve clustering; most clustering algorithms do not allow for the use of an arbitrary function to compare the input data. K-Means for example, which has been used in previous unsupervised feature learning work, is designed to work with the Euclidean distance and no other measure. Modified versions of K-Means have been used to cluster lightcurves using measures like cross-correlation \citep{rebbapragada2009finding}, but these modifications make the algorithm, at least, an order of magnitude slower. An additional disadvantage is that the number of clusters, $K$, has to be specified as input. Affinity Propagation \citep{frey2007clustering} is a clustering algorithm that works with any input data as long as there is a similarity function defined for their comparison, which is exactly our case with the TWED. This algorithm has the additional advantage that it does not need an apriori specification of the number of clusters to find, and it defines a representative exemplar of each clusters. We use this set of exemplars as our lightcurve model.

What follows in this section is a detailed explanation of the Edit Distance for Time Series, followed by a definition of the Time Warp Edit Distance, and lastly a detailed description of the Affinity Propagation clustering algorithm.

\subsection{Edit Distance for Time Series}
\label{sec:edit_distance}

The Levenshtein Distance \citep{levenshtein1966binary}, commonly known as Edit Distance, is a distance metric used in many applications in computer science to assess the similarity between two strings of characters. The Levenshtein Distance (LD) is defined as the smallest number of insertions, deletions and substitutions required to change one string into another. The ideas behind LD have been extended for time series matching. What follows is a brief definition of the matching problem applied to time series. 

Let $U$ be the set of finite time series: $U = \{ X_1^p | p \in \mathbb{N} \},$ $X_1^p$ is a time series with discrete time index between $1$ and $p$. 
Let $x_i$ be the $i$-th sample of time series $X$. We consider that $x_i \in S \times T$ where $S \subset \mathbb{R}$ embeds the time series values and $T \subset \mathbb{R}$ embeds the time variable. We say that $x_i = (m_{x_i}, t_{x_i})$ where $m_{x_i} \in S$ and $t_{x_i} \in T$, with $t_{x_i} > t_{x_j}$ whenever $i > j$ (time stamp strictly increases in the sequence of samples). $X_i^j$ with $i < j$ is the sub time series consisting of the $i$-th through the $j$-th sample (inclusive) of X. $|X|$ denotes the length (the number of samples) of X. $\Lambda$ denotes the null sample.

An edit operation is a pair $(x, y) \neq (\Lambda, \Lambda)$ of time series samples, written $x \rightarrow y$. Time series $Y$ results from the application of the edit operation $x \rightarrow y$ to time series $X$, written $X \Rightarrow Y$ via $x \rightarrow y$, if $X = \sigma x \tau$ and $Y = \sigma y \tau$ for some time series (both time series are the same except for subset $x$ and $y$). We call $x \rightarrow y$ a match operation if $x \neq \Lambda$ and $y \neq \Lambda$, a delete operation if $y = \Lambda$, an insert operation if $x = \Lambda$. Similarly to the edit distance defined for strings, we can define $\delta(X, Y)$ as the similarity between any two time series X and Y of finite lengths $p$ and $q$ as:

\begin{equation*}
\delta(X^p_1, Y^q_1) = \min \begin{cases}
    \delta(X^{p-1}_1, Y^q_1) + \Gamma(x_p \rightarrow \Lambda) & {\rm delete} \\
    \delta(X^{p-1}_1, Y^{q-1}_1) + \Gamma(x_p \rightarrow y_q) & {\rm match}\\
    \delta(X^p_1, Y^{q-1}_1) + \Gamma(\Lambda \rightarrow y_q) & {\rm insert}
\end{cases}
\end{equation*} where $p \geqslant 1, q \geqslant 1$ and $\Gamma$ is an arbitrary cost function which assigns a nonnegative real number $\Gamma(x \rightarrow y)$ to each edit operation $x \rightarrow y$.

It is worth pointing out that in the context of astronomical lightcurves, the notation $X_1^p$ corresponds to a lightcurve with $p$ observations, $x_i = (m_{x_i}, t_{x_i})$ is the $i-th$ observation with $m_{x_i}$ being its photometric magnitude and $t_{x_i}$ the observation time.

\subsection{Time Warp Edit Distance}
\label{sec:twed}

\begin{figure}[h]
\centering
\includegraphics[width=8.5cm]{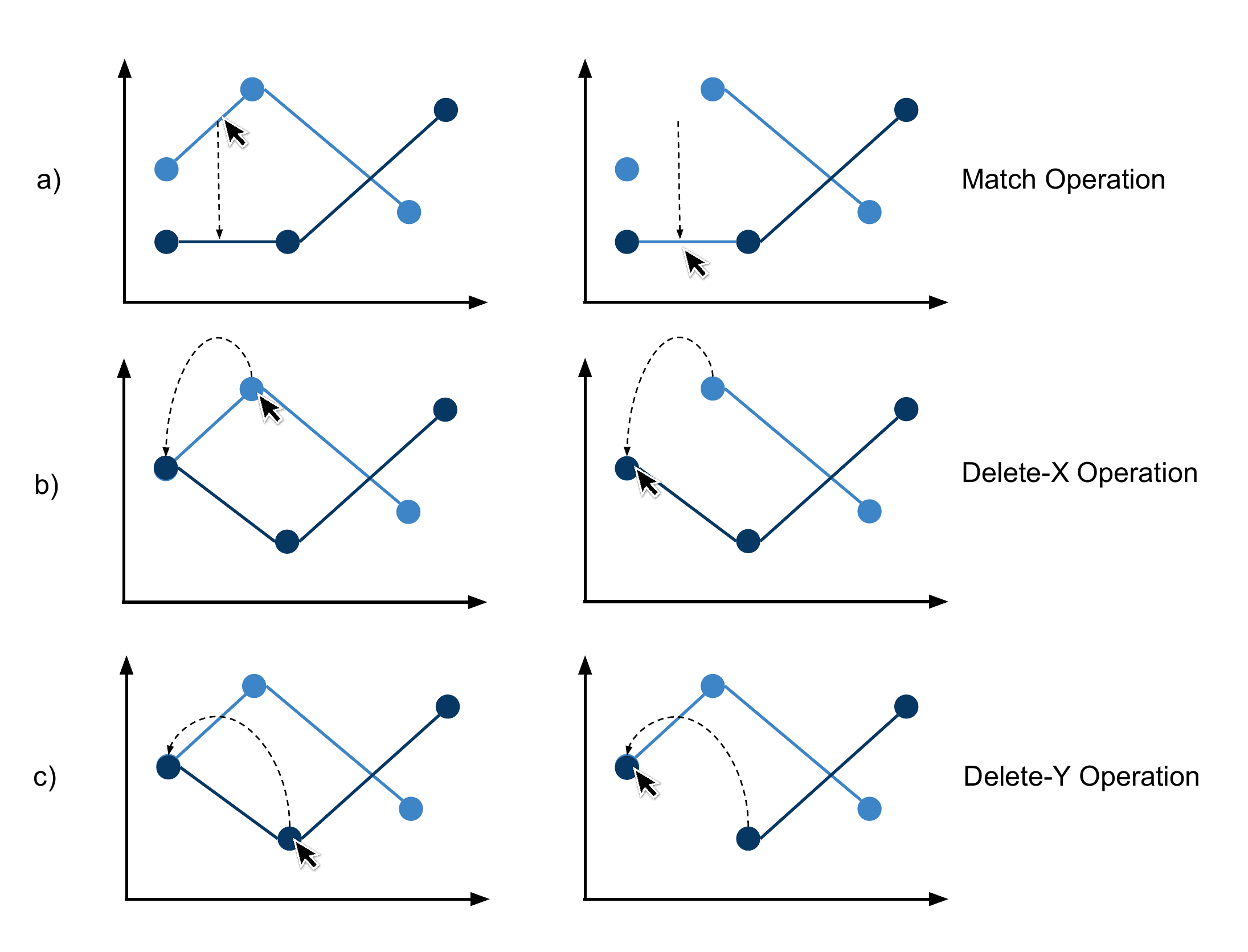}
\caption{Edit operations in a graphical editor. Time series X and Y are depicted in light blue and dark blue, respectively.}
\label{fig:twed}
\end{figure}

Time Warp Edit Distance (TWED) is a similarity measure for time series based on the Edit Distance for time series but aims to provide an elastic metric for time series matching by taking the time differences into account when penalizing edit operations. TWED's edit operations are best understood as tools for superimposing two time series on a 2D graphical editor. Instead of \textit{match}, \textit{delete} and \textit{insert} operations, TWED defines the \textit{match}, \textit{delete-X} and \textit{delete-Y} operations:

\begin{itemize}
\item \textit{match}: The \textit{match} operation (Figure \ref{fig:twed}a) consists of matching a segment $(x_{i-1}, x_i)$ of X with a segment $(y_{j-1}, y_j)$ of Y. In the graphical editor paradigm, the operation consists of clicking on the line which represents segment $(x_{i-1}, x_i)$ and dragging and dropping it onto the line which represents segment $(y_{j-1}, y_j)$. The cost of this operation is proportional to the sum of the distances between corresponding samples of the segments:  $\abs{y_j - x_i}$ and $\abs{y_{j-1} - x_{i-1}}$.

\item \textit{delete-X}: The \textit{delete-X} operation (Figure \ref{fig:twed}b) consists of deleting a sample $x_i$. In the graphical editor paradigm, the operation consists of clicking on the point which represents sample $x_i$ and dragging and dropping it onto the point which represents sample $x_{i-1}$. The cost associated with this delete operation is proportional to the length of the vector $(x_i - x_{i-1})$ to which a constant penalty $\lambda$ is added.
\item \textit{delete-Y}: Just like the previous operation, \textit{delete-Y} operation (Figure \ref{fig:twed}c) consists of deleting a sample $y_i$. In the graphical editor paradigm, the operation consists of clicking on the point which represents sample $y_i$ and dragging and dropping it onto the point which represents sample $y_{i-1}$. The cost associated with this delete operation is proportional to the length of the vector $(y_i - y_{i-1})$ to which a constant penalty $\lambda$ is added.
\end{itemize}

The three edit operations are illustrated in Figure \ref{fig:twed} following the idea of edit operations in a graphical editor paradigm.

The previous operations together with the definitions of section \ref{sec:edit_distance} provide the basis for the definition of TWED:

\begin{equation*}
\delta_{\lambda, \gamma}(X^p_1, Y^q_1) = \min \begin{cases}
    \delta_{\lambda, \gamma}(X^{p-1}_1, Y^q_1) + \Gamma_{\textbf{x}} & {\rm del}-X \\
    \delta_{\lambda, \gamma}(X^{p-1}_1, Y^{q-1}_1) + \Gamma_{\textbf{xy}} & {\rm match}\\
    \delta_{\lambda, \gamma}(X^p_1, Y^{q-1}_1) + \Gamma_{\textbf{y}} & {\rm del}-Y
\end{cases}
\end{equation*} where \begin{align*}
\Gamma_{\textbf{x}} &= \abs{m_{x_p} - m_{x_{p-1}}} + {\gamma}\abs{t_{x_p} - t_{x_{p-1}}} + \lambda \\
\Gamma_{\textbf{xy}} &= \abs{m_{x_p} - m_{y_q}} + {\gamma}\abs{t_{x_p} - t_{y_q}} \\
&\quad{} + \abs{m_{x_{p-1}} - m_{y_{q-1}}} + {\gamma}\abs{t_{x_{p-1}} - t_{y_{q-1}}} \\
\Gamma_{\textbf{y}} &= \abs{m_{y_q} - m_{y_{q-1}}} + {\gamma}\abs{t_{y_q} - t_{y_{q-1}}} + \lambda \end{align*}

It is important to note that parameter $\gamma$ controls the \quotes{elasticity} of TWED: the higher it is, the higher the penalties related to time stamp differences. Like many proposed time series similarity measures, TWED is calculated with a simple dynamic programming algorithm with running time $O(pq)$. The recursion is initialized to $\delta_{\lambda, \gamma}(X^i_1, Y^1_1) = \infty, \forall i > 1$; $\delta_{\lambda, \gamma}(X^1_1, Y^j_1) = \infty, \forall j > 1$ and $\delta_{\lambda, \gamma}(X^1_1, Y^1_1) = 1$.

To better understand TWED, consider the following example where two match operations are performed in total. Let $X_1^3$ and $Y_1^3$ be two time series with three samples each, $X = \{(1, 3), (3, 6), (8, 8)\}$ and $Y = \{(2, 1), (5, 8), (9, 7)\}$. Let $\lambda = \gamma = 1$. We have:

\begin{align*}
\delta_{\lambda, \gamma}(X^3_1, Y^3_1) =\min \begin{cases}
    \delta_{\lambda, \gamma}(X^2_1, Y^3_1) + \Gamma_{\textbf{x}} \\
    \delta_{\lambda, \gamma}(X^2_1, Y^2_1) + \Gamma_{\textbf{xy}} \\
    \delta_{\lambda, \gamma}(X^3_1, Y^2_1) + \Gamma_{\textbf{y}}
\end{cases}
\end{align*} where \begin{align*}
\Gamma_{\textbf{x}} &= \abs{m_{x_3} - m_{x_{2}}} + {\gamma}\abs{t_{x_3} - t_{x_{2}}} + \lambda \\
&= \abs{8 - 6} + 1 \times \abs{8 - 3} + 1 \\
&= 8 \\
\Gamma_{\textbf{xy}} &= \abs{m_{x_3} - m_{y_3}} + {\gamma}\abs{t_{x_3} - t_{y_3}} \\
&\quad{} + \abs{m_{x_2} - m_{y_2}} + {\gamma}\abs{t_{x_2} - t_{y_2}} \\
&= \abs{8 - 7} + 1 \times \abs{8 - 9} \\
&\quad{} + \abs{6 - 8} + 1 \times \abs{3 - 5} \\
&= 6 \\
\Gamma_{\textbf{y}} &= \abs{m_{y_3} - m_{y_2}} + {\gamma}\abs{t_{y_3} - t_{y_2}} + \lambda \\
&= \abs{7 - 8} + 1 \times \abs{9 - 5} + 1 \\
&= 6
\end{align*} so \begin{align*}
\delta_{\lambda, \gamma}(X^3_1, Y^3_1) &= \min \begin{cases}
    \delta_{\lambda, \gamma}(X^2_1, Y^3_1) + 8 & {\rm del}-X \\
    \delta_{\lambda, \gamma}(X^2_1, Y^2_1) + 6 & {\rm match}\\
    \delta_{\lambda, \gamma}(X^3_1, Y^2_1) + 6 & {\rm del}-Y
\end{cases}
\end{align*} we then calculate $\delta_{\lambda, \gamma}(X^2_1, Y^3_1)$, $\delta_{\lambda, \gamma}(X^3_1, Y^2_1)$ and $\delta_{\lambda, \gamma}(X^2_1, Y^2_1)$: \begin{align*}
\delta_{\lambda, \gamma}(X^2_1, Y^2_1) &=\min \begin{cases}
    \delta_{\lambda, \gamma}(X^1_1, Y^2_1) + \Gamma_{\textbf{x}} & {\rm del}-X\\
    \delta_{\lambda, \gamma}(X^1_1, Y^1_1) + \Gamma_{\textbf{xy}} & {\rm match}\\
    \delta_{\lambda, \gamma}(X^2_1, Y^1_1) + \Gamma_{\textbf{y}} & {\rm del}-Y \end{cases}\\
    &= \min \begin{cases}
    \infty \\
    0 + \Gamma_{\textbf{xy}} \\
    \infty \end{cases} \\
\Gamma_{\textbf{xy}} &= \abs{m_{x_2} - m_{y_2}} + {\gamma}\abs{t_{x_2} - t_{y_2}} \\
&\quad{} + \abs{m_{x_1} - m_{y_1}} + {\gamma}\abs{t_{x_1} - t_{y_1}} \\
&= \abs{6 - 8} + 1 \times \abs{3 - 5} \\
&\quad{} + \abs{3 - 1} + 1 \times \abs{1 - 2} \\
&= 7
\end{align*} so \begin{align*}
\delta_{\lambda, \gamma}(X^2_1, Y^3_1) &= \min \begin{cases}
    \delta_{\lambda, \gamma}(X^1_1, Y^3_1) + \Gamma_{\textbf{x}} & {\rm del}-X\\
    \delta_{\lambda, \gamma}(X^1_1, Y^2_1) + \Gamma_{\textbf{xy}} & {\rm match}\\
    \delta_{\lambda, \gamma}(X^2_1, Y^2_1) + \Gamma_{\textbf{y}} & {\rm del}-Y \end{cases}\\
    &= \min \begin{cases}
    \infty \\
    \infty \\
    \delta_{\lambda, \gamma}(X^2_1, Y^2_1) + 6 \end{cases} \\
    &= 13 \\
\delta_{\lambda, \gamma}(X^3_1, Y^2_1) &= \min \begin{cases}
    \delta_{\lambda, \gamma}(X^2_1, Y^2_1) + \Gamma_{\textbf{x}} & {\rm del}-X\\
    \delta_{\lambda, \gamma}(X^2_1, Y^1_1) + \Gamma_{\textbf{xy}} & {\rm match}\\
    \delta_{\lambda, \gamma}(X^3_1, Y^1_1) + \Gamma_{\textbf{y}} & {\rm del}-Y \end{cases}\\
    &= \min \begin{cases}
    \delta_{\lambda, \gamma}(X^2_1, Y^2_1) + 8 \\
    \infty \\
    \infty \end{cases} \\
    &= 15
\end{align*} and finally \begin{align*}
\delta_{\lambda, \gamma}(X^3_1, Y^3_1) &= \min \begin{cases}
    13 + 8 \\
    7 + 6 \\
    15 + 6 \end{cases} \\
&= 13
\end{align*} 

The distance between $X_1^3$ and $Y_1^3$ is 13. If we were to calculate the TWED between two identical time series, the matching cost $\Gamma_{\textbf{xy}}$ would  be zero at each step. Is it easy to see then that the TWED between two identical time series is zero since at each step the match operation of zero cost would be chosen.

\subsection{Affinity Propagation}
\label{sec:affinity}

Affinity Propagation \citep{frey2007clustering} is a clustering algorithm which aims to find representative exemplars from its input data. This algorithm views each data point as a node in a network, and recursively transmits real-valued messages along the edges of the network until a satisfactory set of exemplar points emerges. The magnitude of the transmitted messages reflects the \quotes{affinity} that one data point has for choosing another point as its exemplar.

The algorithm input is a matrix of real-valued similarities between data points, where $s(i, k)$ is the similarity between the data points with indexes $i$ and $k$. A higher value of $s(i,k)$ reflects a higher similarity. This measure is usually set to the negative Euclidean distance (distant points get low similarities), but the method can be applied to any arbitrary similarity measure. The values along the diagonal of the similarity matrix, $s(k, k)$ are called \quotes{preferences}, and a larger value reflects a higher likelihood of being chosen as an exemplar during clustering. When no data point should be favoured during clustering, like in our experiments, $s(k, k)$ should be set to a common value for all $k$. Another significant advantage of this algorithm besides the aforementioned flexitility is that in contrast to other common clustering algorithms like K-Means, Affinity Propagation doesn't require the number of clusters to be specified in advance. The number of clusters (number of exemplars) found is affected by both the values set for preferences and the message passing procedure. In our experiments, we set the preferences to the median similarity between all points, which produces a moderate number of clusters \citep{frey2007clustering}. Another value used for the preferences is the minimum similarity, which produces a small number of clusters.

Data points exchange two different kinds of messages during clustering: \quotes{responsibility} $r(i, k)$ and \quotes{availability} $a(i, k)$. The first reflects the accumulated evidence for how good point $k$ is to serve as an exemplar to point $i$, while the second reflects how appropriate it would be for point $i$ to choose point $k$ as its exemplar. The availabilities are initialized to zero: $a(i, k) = 0$. The responsibilities are then computed using the following update rule: $$ r(i, k) \leftarrow s(i, k) - \max_{k' s.t. k' \neq k} \{a(i, k') + s(i, k')\} $$ This update rule should be seen as a competition between all candidate exemplars for ownership of a data point. The availability update rule, on the other hand, gathers evidence from data points as to wether a candidate exemplar would be a good exemplar: $$ a(i, k) \leftarrow \min \{0, r(k, k) + \sum_{i' s.t. i' \neq \{i, k\}} \max\{0, r(i', k)\} $$ The availability $a(i, k)$ is set to the self-responsibility $r(k, k)$ plus the sum of the positive responsibilities candidate exemplar $k$ receives from other points. Only the positive portions of incoming responsibilities are added, because it is only necessary for a good exemplar to explain some data points well (positive responsibilities), regardless of how how poorly it explains other data points (negative responsibilities). The \quotes{self-availability} $a(k,k)$, is updated with the following rule: $$a(k, k) \leftarrow \sum_{i' s.t. i' \neq k} \max \{0, r(i, k)\} $$ This message reflects accumulated evidence that point $k$ is an exemplar based on the positive responsibilities sent to candidate exemplar $k$ from other points.

At any moment during affinity propagation, availabilities and responsibilities can be combined to identify exemplars. For point $i$, the value of $k$ that maximizes $a(i, k) + r(i, k)$ either identifies point $i$ as an exemplar if $k = i$, or identifies the data point that is the exemplar for point $i$. The message-passing procedure may be terminated after a fixed number of iterations, after changes in the messages fall below a threshold, or after the local decisions stay constant for some number of iterations.

\section{Method Description}
\label{sec:method}

Our method draws from what was proposed in \citet{coates2012learning} for the domain of images, with substantial modifications to make our new algorithm work well with lightcurves. As \citet{keogh2005clustering} demonstrated, time series subsequence clustering with K-Means and Euclidean distance will very seldom produce meaningful results. Furthermore, the Euclidean distance is not well defined for the comparison of two lightcurves since the two time-series will rarely have the same length because they are not evenly sampled. To overcome this problem, we employ the  Time Warp Edit Distance (section \ref{sec:twed}) together with an appropriate clustering algorithm that works well with any similarity measure for its data, Affinity Propagation (section \ref{sec:affinity}).

\begin{figure*}
\centering
\includegraphics[width=17cm]{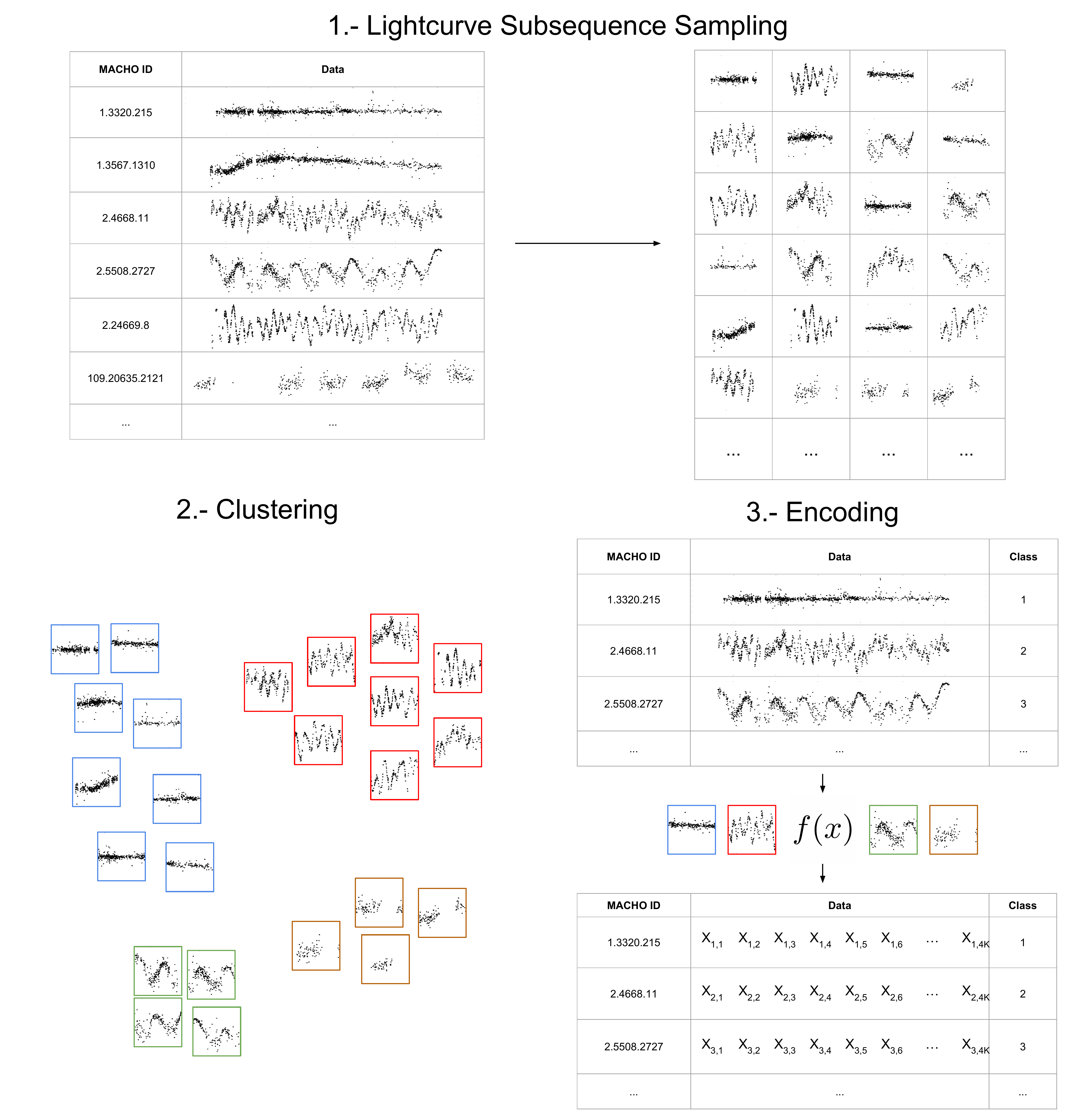}
\caption{Method overview illustration: In the first step, we draw random subsequences from lightcurves to form a large set of lightcurve fragments. The second step consists of clustering these fragments with the Affinity Propagation algorithm. The third step consists of using the representative exemplars found during clustering to encode a training set of labeled lightcurves to a new representation more suitable for automatic classification tasks.}
\label{fig:overview}
\end{figure*}

Our algorithm consists of three main steps. In the first step, we randomly sample subsequences from lightcurves to form a large set of lightcurve fragments. The second step consists of clustering these fragments with the Affinity Propagation algorithm and the Time Warp Edit Distance similarity measure, both described in detail in sections \ref{sec:affinity} and \ref{sec:twed} respectively. The third step consists of using the representative exemplars, found during clustering, to encode a training set of labeled lightcurves to a new representation  for  the classification tasks. Figure \ref{fig:overview} provides an illustrated overview of the process.

\subsection{Lightcurve Subsequence Sampling}
\label{sec:sampling}

\begin{figure}[h]
\centering
\includegraphics[width=8.5cm]{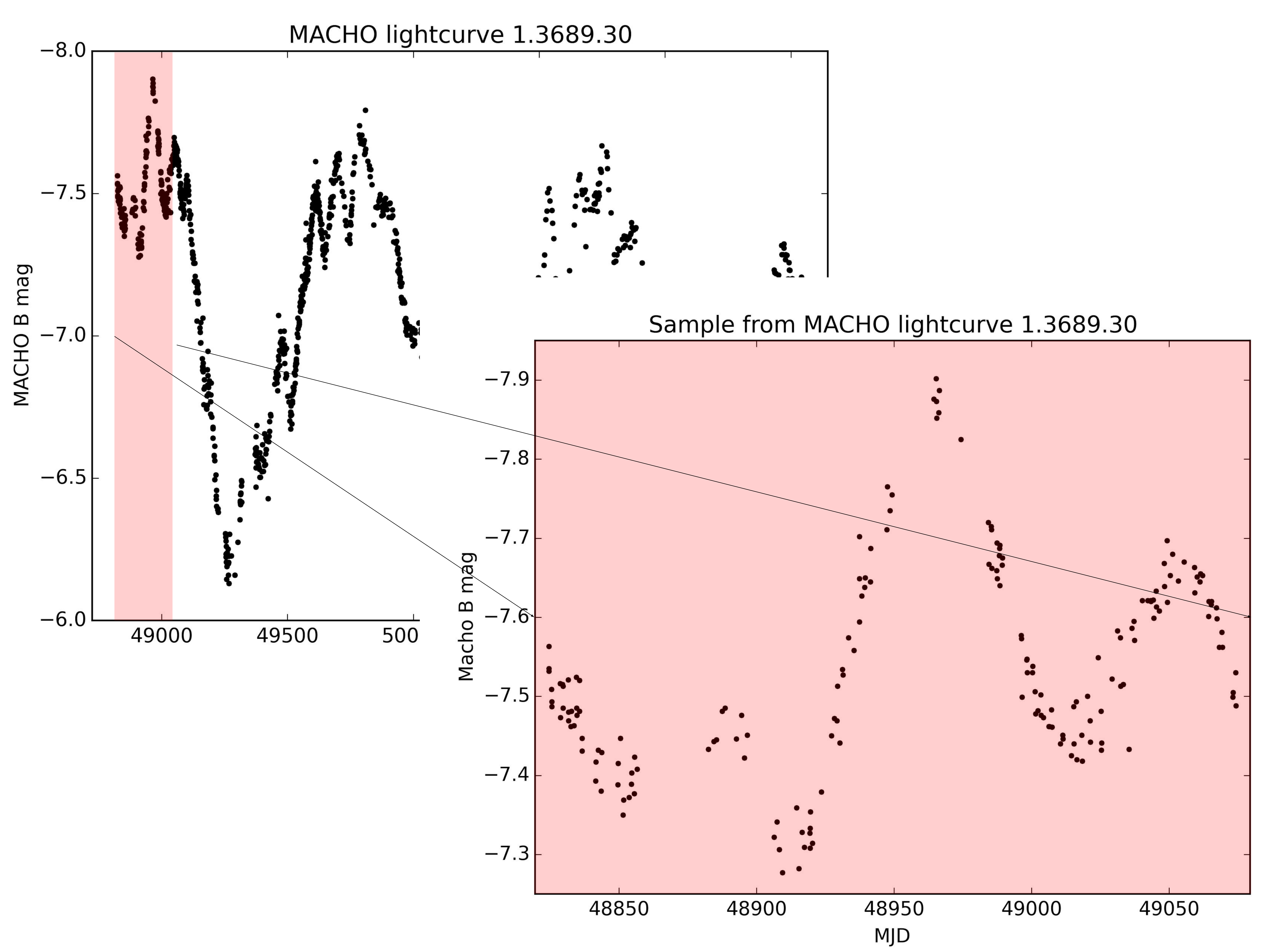}
\caption{Lightcurve subsequence sampling. We sample a subsequence of the lightcurve by extracting all the observations in a given time window (translucent red), $t_w$.}
\label{fig:lc_sampling}
\end{figure}

To get the data that we want to cluster, we randomly sample $N$ subsequences of lightcurves from a given dataset by extracting all the observations in a given time window, $t_w$. The idea behind sampling small time windows and not using the whole lightcurve is to force our model to capture local patterns in the data. This procedure is illustrated in Figure \ref{fig:lc_sampling}.

\subsection{Affinity Propagation Clustering}
\label{sec:clustering}

\begin{figure}[h]
\centering
\includegraphics[width=8.5cm]{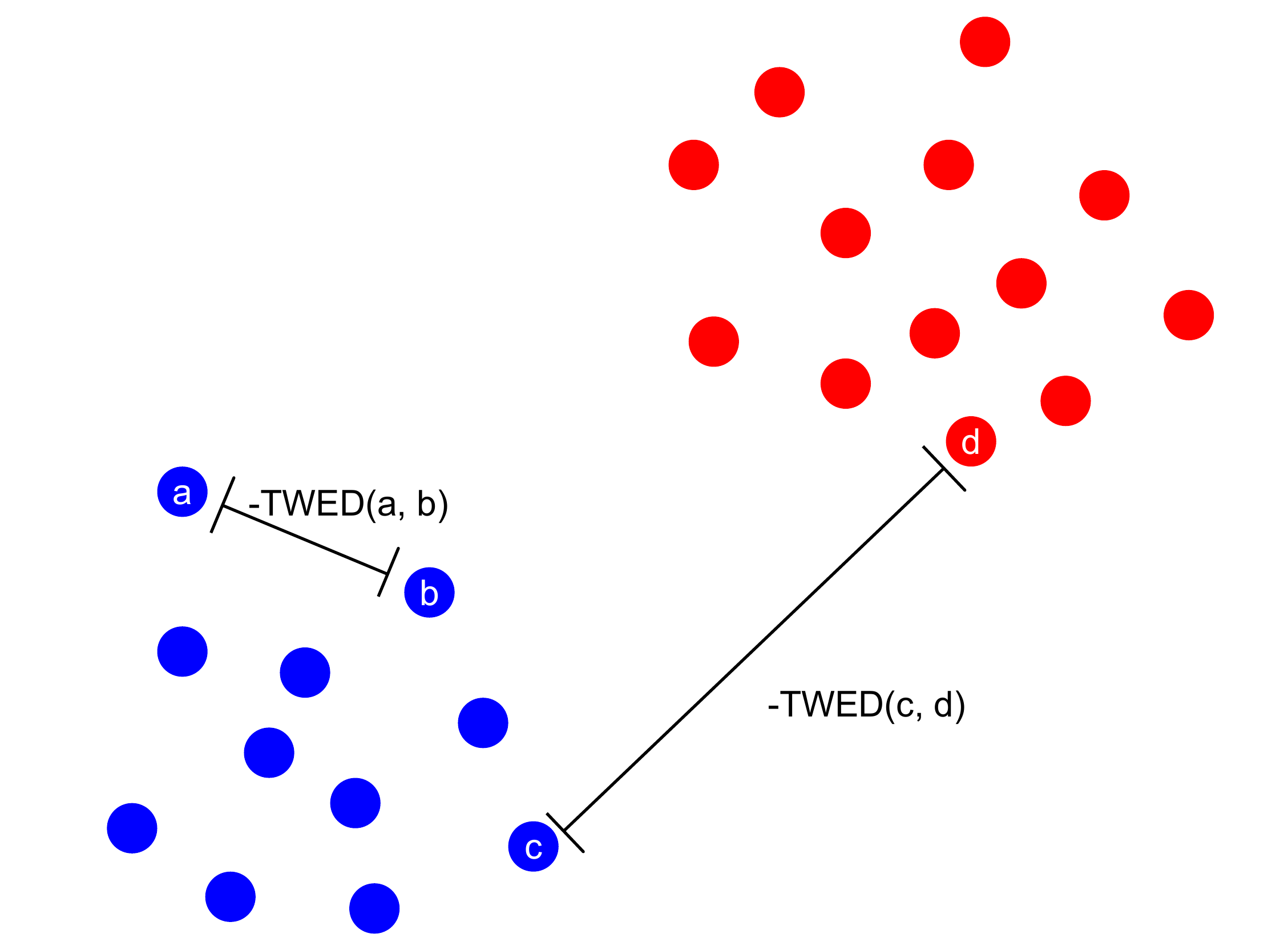}
\caption{Lightcurve clustering. The lightcurve subsequences (represented by the colored dots) are grouped into clusters according to their affinity measure, which in this case is the negative TWED.}
\label{fig:clustering}
\end{figure}

After collecting $N$ lightcurve fragments from our data, we run the Affinity Propagation clustering algorithm with the set of fragments extracted in the first step as input data to find a set of representative lightcurve subsequences. This process is illustrated in Figure \ref{fig:clustering}. The affinity measure used during clustering is the negative TWED: $ -\delta_{\lambda, \gamma} (X_1^p, Y_1^q) $, where $X_1^p$ and $Y_1^q$ are two lightcurve fragments. We use the negative TWED since that way a greater distance means a lesser degree of similarity. After the clustering is completed, we have a set of $K$ representative exemplars from the data, which capture common local patterns occurring in the time series.

\subsection{New Representation}
\label{sec:encoding}

With the $K$ exemplars found during the clustering step, we use a feature mapping function $f$ to map any lightcurve fragment to a new feature space. The idea is to encode any lightcurve fragment as a $K$-dimensional vector where each index of the vector will represent a degree of similarity between the lightcurve fragment and each of the $K$ exemplars. Our choice of $f$ is: 
$$ f_k = max\{0, \mu(\delta_{\lambda, \gamma}) - \delta_{\lambda, \gamma}(X_1^p, c^{(k)}) \} $$
 where $X_1^p$ is a lightcurve fragment with $p$ observations, $ c^{(k)}$ is the $k$-th exemplar, $\mu(\delta_{\lambda, \gamma})$ is the average TWED between the fragment and all the other exemplars. This means that the value of any given index of the vector will be 0 if the distance to that exemplar is above average, and a positive value when the distance is below the average. This value is larger when the fragment is more similar to the exemplar. It is expected that roughly half the values in any given vector will be zero, which is a favorable condition for our classification procedure, detailed in section \ref{sec:classification}.

\begin{figure*}
\centering
\includegraphics[width=17cm]{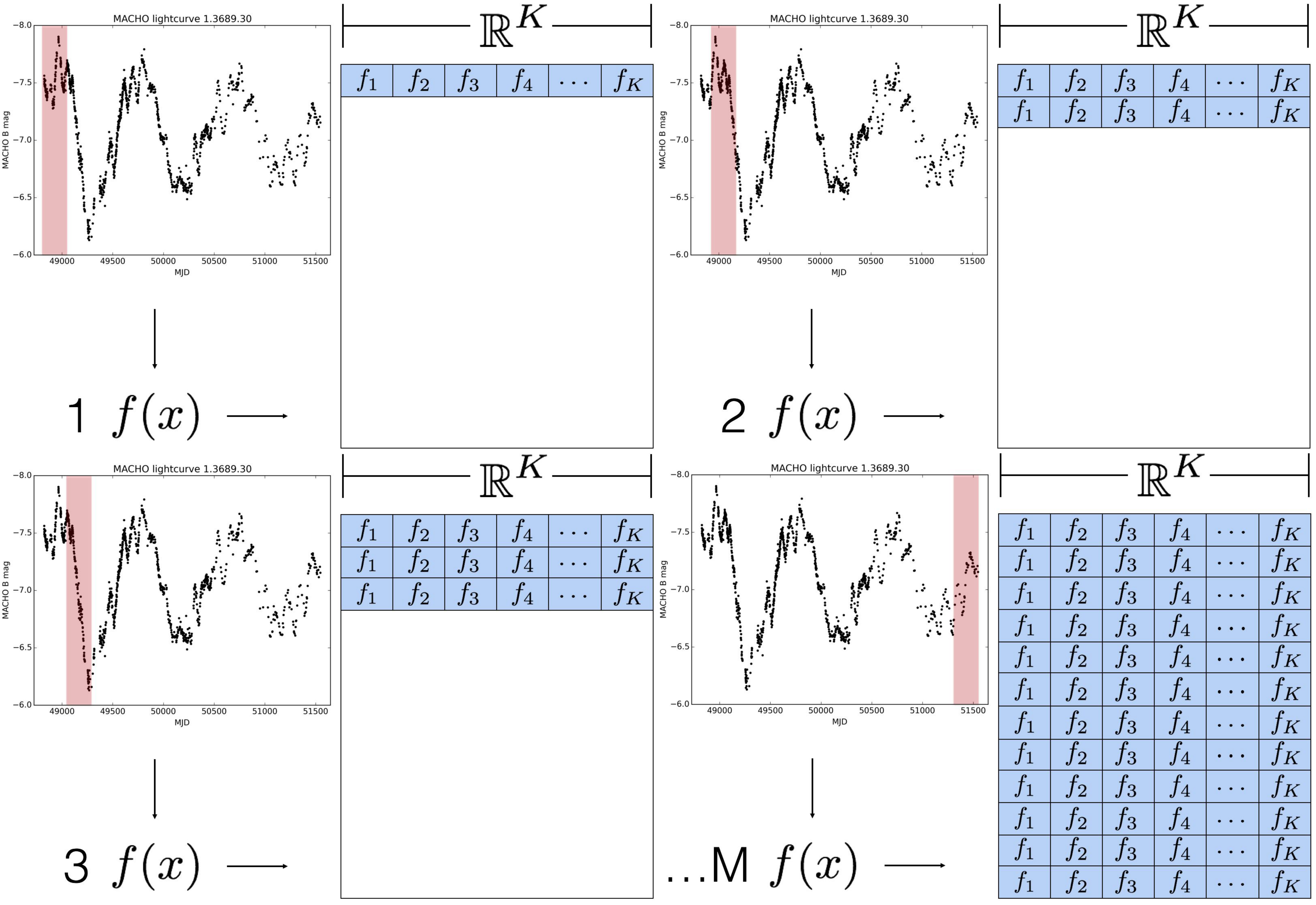}
\caption{Sliding window process. The sliding window (translucent red) extracts a subsequence of the lightcurve at each step, which is encoded as a $K$-dimensional vector by our encoding function $f$. The window moves sequentially along the time-axis, extracting and encoding one subsequence at each step.}
\label{fig:sliding_window}
\end{figure*}

Given this feature mapping function, we can now encode a complete lightcurve in our new representation by applying $f$ to sequential fragments of the lightcurve. Specifically, given time step $t_s$ and the time window $t_w$, the adjacent fragments are obtained by getting all the time series data in one window and then moving the time window by $t_s$, sliding the window across the whole lightcurve. It is worth noting $t_s$ is usually much smaller than $t_w$, so the extracted fragments overlap significantly. We extract adjacent fragments from each lightcurve until the sliding window reaches the end of the observations; this means that the number of fragments extracted is variable and depends on the length of the lightcurve. If $M$ is the number of fragments extracted from a lightcurve, the final representation is of dimensions $ \mathbb{R}^{M \times K} $. This process is illustrated in Figure~\ref{fig:sliding_window}.

This intermediate representation of a lightcurve is too large for use as direct input to any classification algorithm. To reduce the dimensionality of data while maintaining the maximum amount of information, it is a common practice to perform a procedure called feature pooling \citep{boureau2010theoretical}. Pooling works by aggregating features extracted from a group of adjacent lightcurve fragments. Encoded fragments from windows that are adjacent or relatively close are also very similar, so finding a way to aggregate those features makes sense to reduce the dimensionality of data. In our experiments, we divide the final representation into four equal sized regions and aggregate the features inside each one. For each of the $K$ features, we take the maximum value in each region, a procedure that is called max-pooling. 

The final pooled representation of a complete lightcurve is a vector of size $4 \times K$, significantly smaller than the representation of size $M \times K$, which is obtained after the sliding window step. The number of regions over which to pool the data represents a trade-off between information preservation and dimensionality of the final representation. We chose four as the number of pools that would allow our representation to preserve the maximum amount of information while still maintaining a manageable dimensionality for the classification stage. Empirically, we found 4 to work better in the classification task.

\subsection{Classification}
\label{sec:classification}

The final training set is composed of all of the lightcurves encoded in our new representation together with their original labels. We use this dataset to train a linear Support Vector Machine (SVM) classifier \citep{boser1992training, Cortes:1995}. The Support Vector Machine is a classifier that tries to fit hyperplanes to data to separate classes. For an overview and discussion see \citet{Kim:2012} and references therein.

\section{Data}
\label{sec:data}

The photometric data used in our experiments belongs to two different catalogs, MACHO and OGLE.

\subsection{MACHO Catalog}

The Massive Compact Halo Object (MACHO) is a survey which observed the sky starting in July 1992 and ending in 1999 to detect microlensing events produced by Milky Way halo objects. Several tens of millions of stars where observed in the Large Magellanic Cloud (LMC), Small Magellanic Cloud (SMC) and Galactic bulge \citep{alcock1997macho}.

\subsection{OGLE-III Catalog of Variable Stars}

The Optical Gravitational Lensing Experiment (OGLE) is a wide-field sky survey originally designed to search for microlensing events \citep{paczynski1986gravitational}. The brightness of more than 200 million stars in the Magellanic Clouds and the Galactic bulge is regularly monitored in the time scale of years. A by-product of these observations is an enourmous database of photometric measurements. The OGLE-III Catalog of Variable Stars \citep{Udalski2008AcA} corresponds to the photometric data collected during the third phase of this survey which began in 2001.

\subsection{Training Sets}

For our encoding and classification experiments we used subsets of both MACHO and OGLE surveys, corresponding to sets of labeled photometric data. The MACHO training set is composed of 4835 labeled observations \citep{Kim:2011}. The OGLE training is composed of 5358 labeled variable objects from the OGLE-III Catalog of Variable Stars \citep{Udalski2008AcA}, the per-class composition of both training sets is detailed in tables \ref{table:macho_training} and \ref{table:ogle_training}. The OGLE training set was chosen as a subset of the most represented variable star classes in the catalog with the objective of creating a training set of comparable size to the MACHO dataset.

\begin{table}[H]
\caption{\label{table:macho_training}MACHO Training Set Composition}
\centering
\begin{tabular}{ c | c | c |}  
\cline{2-3}
{} & Class & Number of Objects \\
\hline
\multicolumn{1}{ |c| }{1} & Non Variable & $3613$\\
\multicolumn{1}{ |c| }{2} & Quasar & $17$\\
\multicolumn{1}{ |c| }{3} & Be Star & $55$\\
\multicolumn{1}{ |c| }{4} & Cepheid & $103$\\
\multicolumn{1}{ |c| }{5} & RR Lyrae & $551$\\
\multicolumn{1}{ |c| }{6} & Eclipsing Binary & $42$\\
\multicolumn{1}{ |c| }{7} & MicroLensing & $173$\\
\multicolumn{1}{ |c| }{8} & Long Period Variable & $281$\\
\hline

\end{tabular}
\end{table}

\begin{table}[H]
\caption{\label{table:ogle_training}OGLE-III Training Set Composition.}
\centering
\begin{tabular}{ c | c | c |}  
\cline{2-3}
{} & Class & Number of Objects \\
\hline
\multicolumn{1}{ |c| }{1} & Cepheid & $992$\\
\multicolumn{1}{ |c| }{2} & Type 2 Cepheid & $476$\\
\multicolumn{1}{ |c| }{3} & RR Lyrae & $971$\\
\multicolumn{1}{ |c| }{4} & Eclipsing Binary & $982$\\
\multicolumn{1}{ |c| }{5} & Delta Scuti & $980$\\
\multicolumn{1}{ |c| }{6} & Long Period Variable & $957$\\
\hline

\end{tabular}
\end{table}

\section{Implementation}
\label{sec:implementation}

Our implementation uses minimal pre-processing: all lightcurves are adjusted to have zero mean and unit variance. For our lightcurve subsequence sampling step (Section \ref{sec:sampling}) we sampled from thousands of unlabeled lightcurves. The parameters we used in our experiments are detailed in Table \ref{table:parameters}. The code for our experiments is available at \url{https://github.com/cmackenziek/tsfl}.

We used the Affinity Propagation and SVM implementations available in the scikit-learn machine learning library \citep{scikit-learn}. We also used the numpy, scipy and pandas libraries for data manipulation and efficient numerical computation \citep{numpy-scipy, mckinney2010data}.

\begin{table*}[]
\centering
\caption{\label{table:parameters}Relevant parameter values.}  
\begin{tabular}{|c|c|c|p{11cm}|}
\hline
  Name & Symbol & Value & Comments  \\  
\hline 
  Time Window & $t_w$ & 250 days & We used 250 days to capture local patterns in the time series while allowing patterns from lightcurves with longer periodicities to be also captured (see Figure \ref{fig:exemplar_plots}). We also considered using the Autocorrelation function length \citep{Kim:2011} but the values of this feature for each class were too different to choose a good common value for all the data.\\
  Time Step & $t_s$ & 10 days & Encoding will work best with as much overlap as possible between the adjacent lightcurve subsequences during the sliding window process (Section \ref{sec:encoding}). Any redundant data will be eliminated through pooling, while no relevant patterns will be missed.\\
  Number of Samples & N & 20,000 & The number of samples affects significantly the performance of the clustering step. We minimised the number of samples subject to still maintaining good classification performance. We consider 20,000 to be a sufficiently large number of samples while still maintaining computational time within reasonable bounds and allowing us to maintain our classification performance.\\
  TWED Elasticity Cost & $\gamma$ & 1e-5 & We chose a relatively low penalty for this parameter to allow for higher \quotes{elasticity} when comparing lightcurve subsequences, in comparison to the values used in \citet{marteau2009time}. \\
  TWED Deletion Cost & $\lambda$ & 0.5 & We chose a mid-point penalty for this parameter so as not to bias the TWED towards matching operations when comparing lightcurve subsequences, in comparison to the values used in \citet{marteau2009time}. \\
  Number of Pooling Regions & - & 4 & The number of regions over which to pool the data represents a trade-off between information preservation and dimensionality of the final representation. We chose 4 as the number of pools that would allow our representation to preserve the maximum amount of information while still maintaining a manageable dimensionality for the classification stage, which is the most common number used throughout the literature \citep{boureau2010theoretical, coates2012learning}. Empirically, we found 4 to work better in the classification stage.\\
\hline
\end{tabular}
\end{table*}

\section{Experimental Results}
\label{sec:results}

In this section, we present the results obtained in our experiments. First, we present the results of the clustering step of our method, which we hope will help the reader gain a qualitative intuition of the inner workings of our algorithm. Then, we present the classification results on all the training sets described in section \ref{sec:data} using two different classifiers and two methods of lightcurve representation: the classical expert-designed time series features and our learned features. Finally, we present an analysis of the classification relevance in terms  of both types of features.

\subsection{Clustering Results}

\begin{figure*}
\centering
\includegraphics[width=17cm]{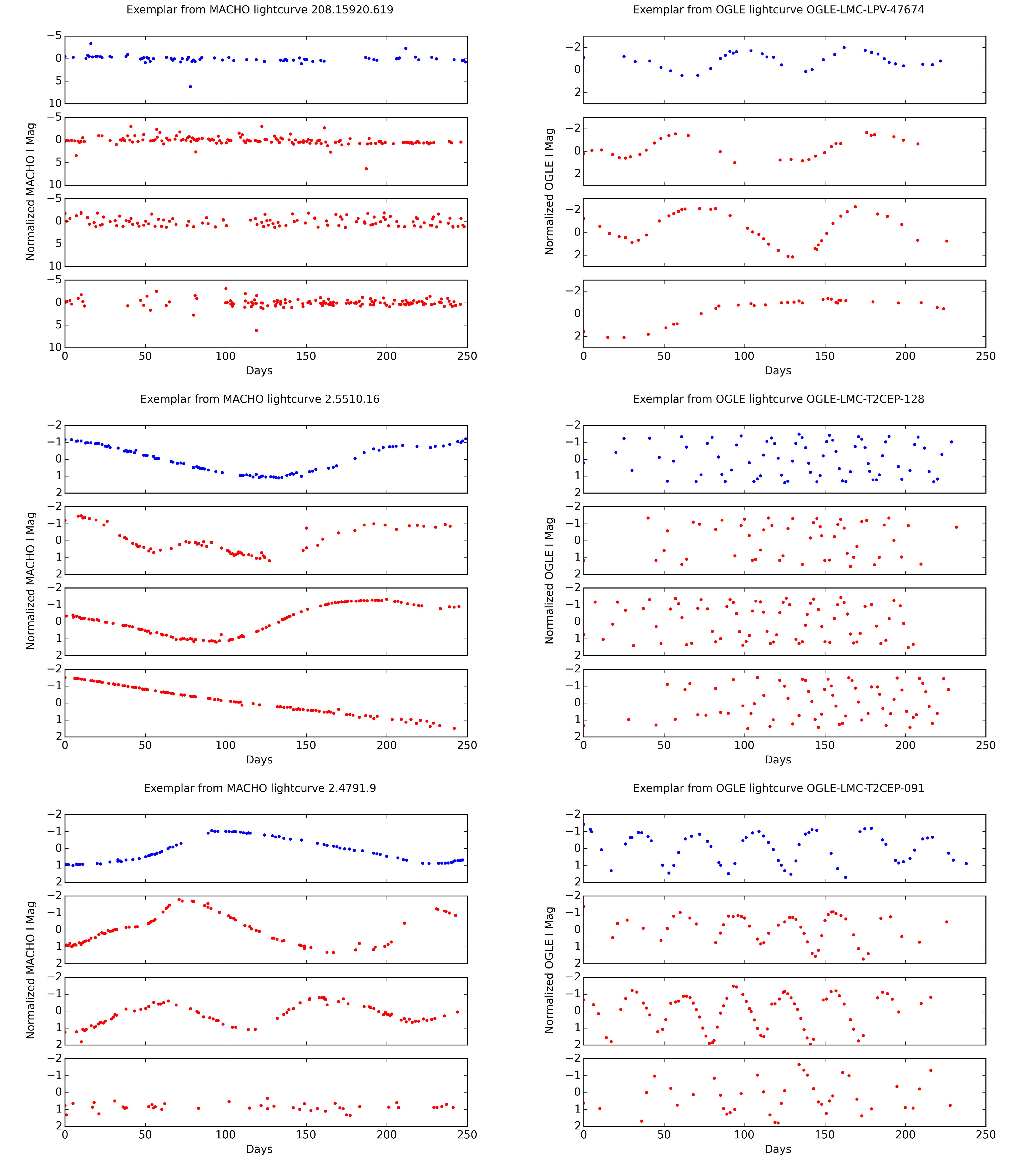}
\caption{Cluster exemplars with members. Exemplars are lightcurve subsequences chosen by the clustering algorithm as the best representatives of their clusters. Each of the six plots shows an exemplar (plotted in blue) together with three other cluster members (plotted in red). We can appreciate how the clustering algorithm successfully groups similar lightcurve subsequences together.}
\label{fig:exemplar_plots}
\end{figure*}

Given that clustering aims to find groups of similar data, one would expect that clustering lightcurve subsequences would group similar patterns in the photometric data. Our results show that this is indeed the case. To show the results of the lightcurve subsequence clustering step described in section \ref{sec:clustering}, we provide plots of some of the learned exemplars together with some other lightcurve subsequences that are members of the same clusters. We can see some of the results in Figure \ref{fig:exemplar_plots}: cluster exemplars are plotted in red together with some members of their respective clusters plotted in blue. We can see that the algorithm captures groups of similar subsequences together. Lightcurve subsequences are grouped by important traits like variability and periodicity. This information is usually estimated with traditional features; our model can automatically group the lightcurve fragments without previously defining what the important criteria are. This previous fact explains how the final encoding step will output relevant data that allows classifiers to distinguish correctly between lightcurves of different classes: subsequences that exhibit different local photometric patterns will be encoded differently since they will be similar to a different subset of exemplars.

\subsection{Training Set Classification Results}

\begin{table*}[]
\centering
\caption{\label{table:macho_results}Classification F-Score on the MACHO training set.}
\begin{tabular}{ c | c | c | c | c|}  
\cline{2-5}
{} & Class & SVM trained with LF & RF trained with TSF & SVM trained with TSF \\
\hline
\multicolumn{1}{ |c| }{1} & Non Variable & 0.991 & 0.991 & 0.875\\
\multicolumn{1}{ |c| }{2} & Quasar & 0.296 & 0.533 & 0.217\\
\multicolumn{1}{ |c| }{3} & Be Star & 0.717 & 0.788 & 0.625\\
\multicolumn{1}{ |c| }{4} & Cepheid & 0.871 & 0.917 & 0.936\\
\multicolumn{1}{ |c| }{5} & RR Lyrae & 0.953 & 0.969 & 0.797\\
\multicolumn{1}{ |c| }{6} & Eclipsing Binary & 0.780 & 0.763 & 0.725\\
\multicolumn{1}{ |c| }{7} & MicroLensing & 0.980 & 0.974 & 0.468\\
\multicolumn{1}{ |c| }{8} & Long Period Variable & 0.975 & 0.947 & 0.802\\
\hline
{} & Weighted Average & 0.975 & 0.978 & 0.807 \\
\cline{2-5}
\end{tabular}
\end{table*}

\begin{table*}[]
\caption{\label{table:ogle_results}Classification F-Score on the OGLE-III training set.}
\centering
\begin{tabular}{ c | c | c | c | c |}  
\cline{2-5}
{} & Class & SVM trained with LF & RF trained with TSF & SVM trained with TSF \\
\hline
\multicolumn{1}{ |c| }{1} & Cepheid & 0.835 & 0.737 & 0.555\\
\multicolumn{1}{ |c| }{2} & Type 2 Cepheid & 0.651 & 0.567 & 0.467\\
\multicolumn{1}{ |c| }{3} & RR Lyrae & 0.749 & 0.868 & 0.649\\
\multicolumn{1}{ |c| }{4} & Eclipsing Binary & 0.862 & 0.602 & 0.458\\
\multicolumn{1}{ |c| }{5} & Delta Scuti & 0.817 & 0.656 & 0.656\\
\multicolumn{1}{ |c| }{6} & Long Period Variable & 0.821 & 0.648 & 0.407\\
\hline
{} & Weighted Average & 0.821 & 0.696 & 0.696\\
\cline{2-5}
\end{tabular}
\end{table*}

To evaluate the classification performance of a classifier trained on our learned features, we must obtain a benchmark with which to compare it. The logical benchmark for this task is the classification performance of a classifier using traditional time series features as input on the same training sets. Classifier performance is measured with a 10-fold stratified cross-validation F-Score on each of the lightcurve classes of a given training set (or test set). Since the data produced by our feature learning method is high dimensional and relatively sparse (each vector will have many zeroes by design, because of our encoding function $f$), we use a Support Vector Machine \citep{Cortes:1995} with a linear kernel as the classifier. To build the time series features training sets, we applied the FATS Library \citep{nun2015fats} which has an exhaustive collection of time series features used throughout the literature. Traditionally, the classifier of choice for lightcurve datasets with time series features has been the Random Forest classifier \citep{Breiman2001}; hence, we decided to compare our SVM with learned features against a Random Forest with the time series features. We also compared our SVM trained on learned features against an SVM trained on time series features. Tables \ref{table:macho_results} and \ref{table:ogle_results} show the results for each training set. The acronym TSF refers to Time Series Features, which are expert-designed features available in the FATS Library and LF refers to Learned Features, which are the features we learn with our method. The SVM classifier performs as well as the Random Forest on the MACHO training set on many classes. Quasars are the only class where the SVM does not achieve comparable performance. We believe this to be due to the relatively low frequency of Quasars in the whole training set, which is known to affect SVM classification performance. On the OGLE-III training set SVM trained on learned features achieved superior results, only performing worst in one class. The first column details the variability class; the second column shows the result of an SVM classifier on  10-fold cross-validation with a linear SVM of both training sets. The learned features achieve a better overall classification performance than the time series features. All weighted averages are calculated using the relative frequency of each class of variable stars in the whole training set.

\subsection{MACHO Field 77 Classification Results}

\begin{table}[H]
\caption{\label{table:f77results}Number of candidates per class on MACHO field 77.}
\centering
\begin{tabular}{| c | c |}
\hline
Class & Number of candidates \\
\hline
Non Variable & 382,306 \\
Quasar & 176 \\
Be Star & 975 \\
Cepheid & 1,459 \\
RR Lyrae & 13,544 \\
Eclipsing Binary & 85,099 \\
MicroLensing & 26,231 \\
Long Period Variable & 1,486 \\
\hline
\end{tabular}
\end{table}

\begin{figure}[h]
\centering
\includegraphics[width=8.5cm]{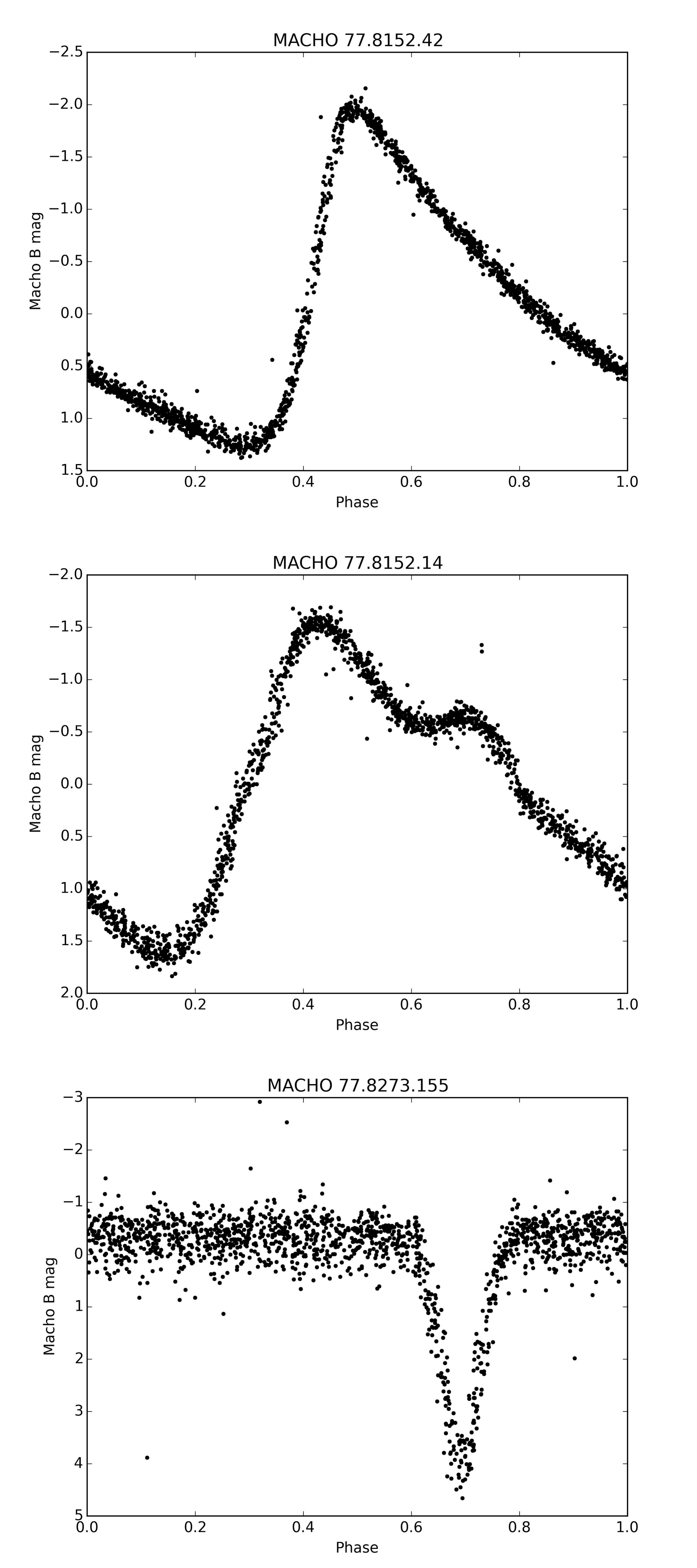}
\caption{New variable star candidate examples. The lightcurves in the plots have been folded since they correspond to periodic stars. The first two lightcurves were classified as Cepheid while the third was classified as an Eclipsing Binary.}
\label{fig:f77lcs}
\end{figure}

In order to discover new variable star candidates, we classified 511,276 lightcurves from field 77 of the MACHO catalog. We found 128,970 variable star candidates, the per-class classification details are shown on Table \ref{table:f77results}. We cross-matched our variable star candidates with the SIMBAD Astronomical Database \citep{wenger2000simbad} to filter out known candidates and found that 15,907 were already known and thus 113,873 are new. Figure \ref{fig:f77lcs} shows examples of our new candidates: the first two lightcurves were classified as Cepheid while the third was classified as an Eclipsing Binary. Our table of candidates is available for download at \url{https://www.dropbox.com/s/fpsktd8aflelp7q/field77results_filtered.csv?dl=0}. We will upload the catalog of our candidates to SIMBAD.

\subsection{Feature Importance}
A very important aspect of a successful classification model is representing the data with relevant features that help the model distinguish between the different labeled data. With this in mind, it would be interesting to analyze how each of the features in a dataset contribute to the final classification task. We performed a feature importance analysis on our learned features and time series features, using the hyperplane coefficients of an SVM with a linear kernel. The general idea behind using the hyperplane coefficients of an SVM is that the importance of a feature in separating between classes is proportional to the magnitude of its corresponding coefficient. A feature that completely separates the classes will have a coefficient of -1 or 1 while a completely irrelevant feature will have a coefficient of 0). A more detailed explanation of the theory behind this analysis can be found in \citep{guyon2002gene} who used SVM coefficients for gene subset selection.

\begin{figure}[h]
\centering
\includegraphics[width=8.5cm]{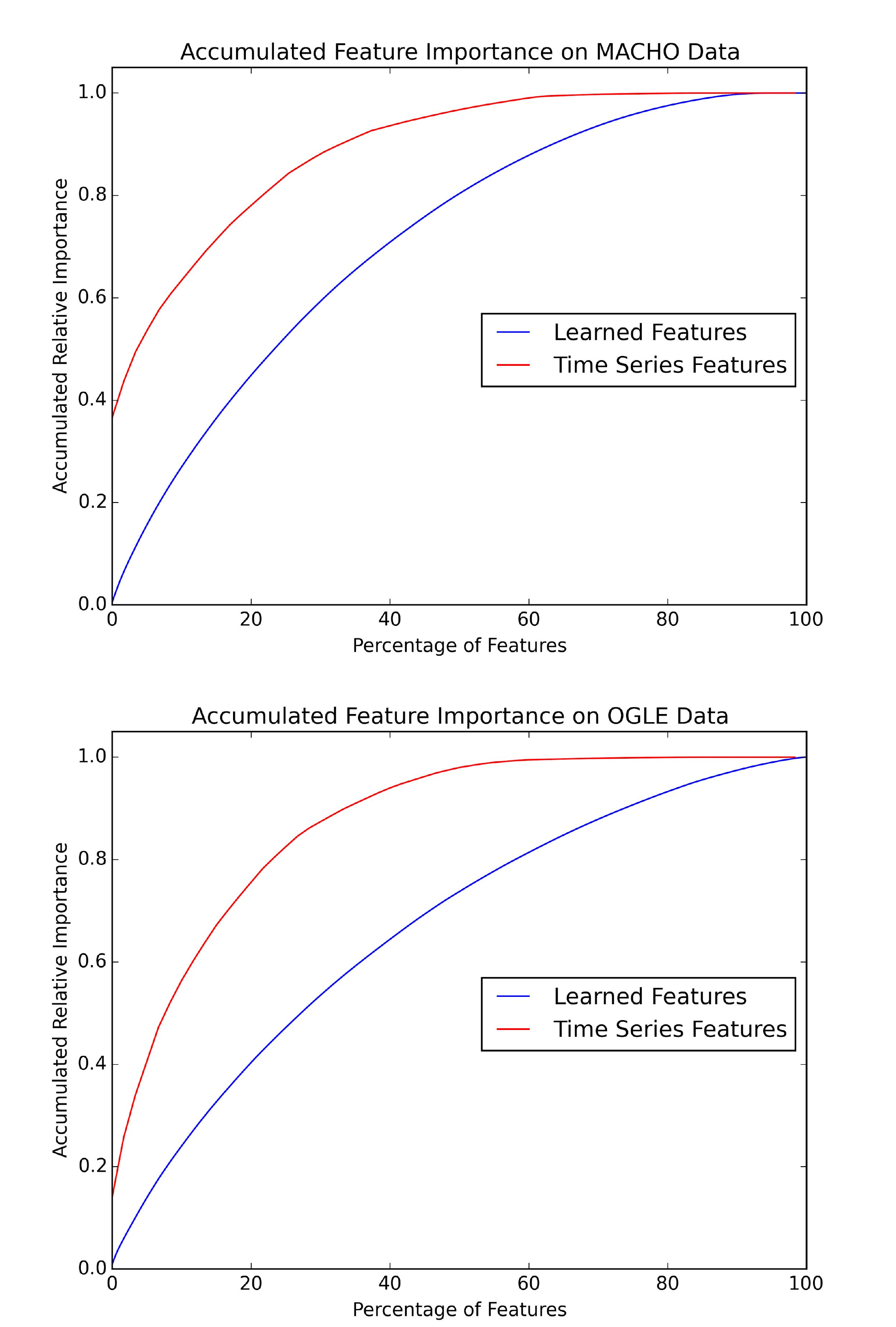}
\caption{Relative importance cumulative sum. A feature set which has lots of features that don't contribute much to classification (i.e. rarely get used by the SVM to separate between classes) would result in a plot where the cumulative sum reaches 1 with a lower percentage of features than a feature set where most of the features are relevant to the classification task for at least some of the classes.}
\label{fig:feature_importance}
\end{figure}

Since our classification problems are multi-class, we get one separating hyperplane for each class. We defined the relative feature importance, $i$, as the sum of the absolute values of each of the feature's coefficients in each hyperplane, divided by the sum of the importance of all features: $$ i_k = \frac{\sum_{j=1}^N |w_k^j|}{\sum_{k=1}^F \sum_{j=1}^N |w_k^j|}  $$ where $w_k^j$ is the coefficient with index $k$ of hyperplane $j$, $N$ is the number of classes and $F$ the number of features. We can visualize at a high level how the two types of features contribute to classification by looking at Figure \ref{fig:feature_importance} which plots the cumulative sum of the relative feature importances versus the percentage of features being added. A feature set that has lots of features that don not contribute much to classification (i.e. rarely get used by the SVM to separate between classes) would result in a plot where the cumulative sum reaches 1 with a low percentage of the features. On the other hand,   a feature set where most of the features are relevant to the classification task for at least some of the classes would result in a plot where the cumulative sum reaches 1 with a high percentage of the features. This is the case of Figure \ref{fig:feature_importance}, where it is evident that a great percentage of the time series features don't contribute much to classification.

\begin{figure*}
\centering
\includegraphics[width=17cm]{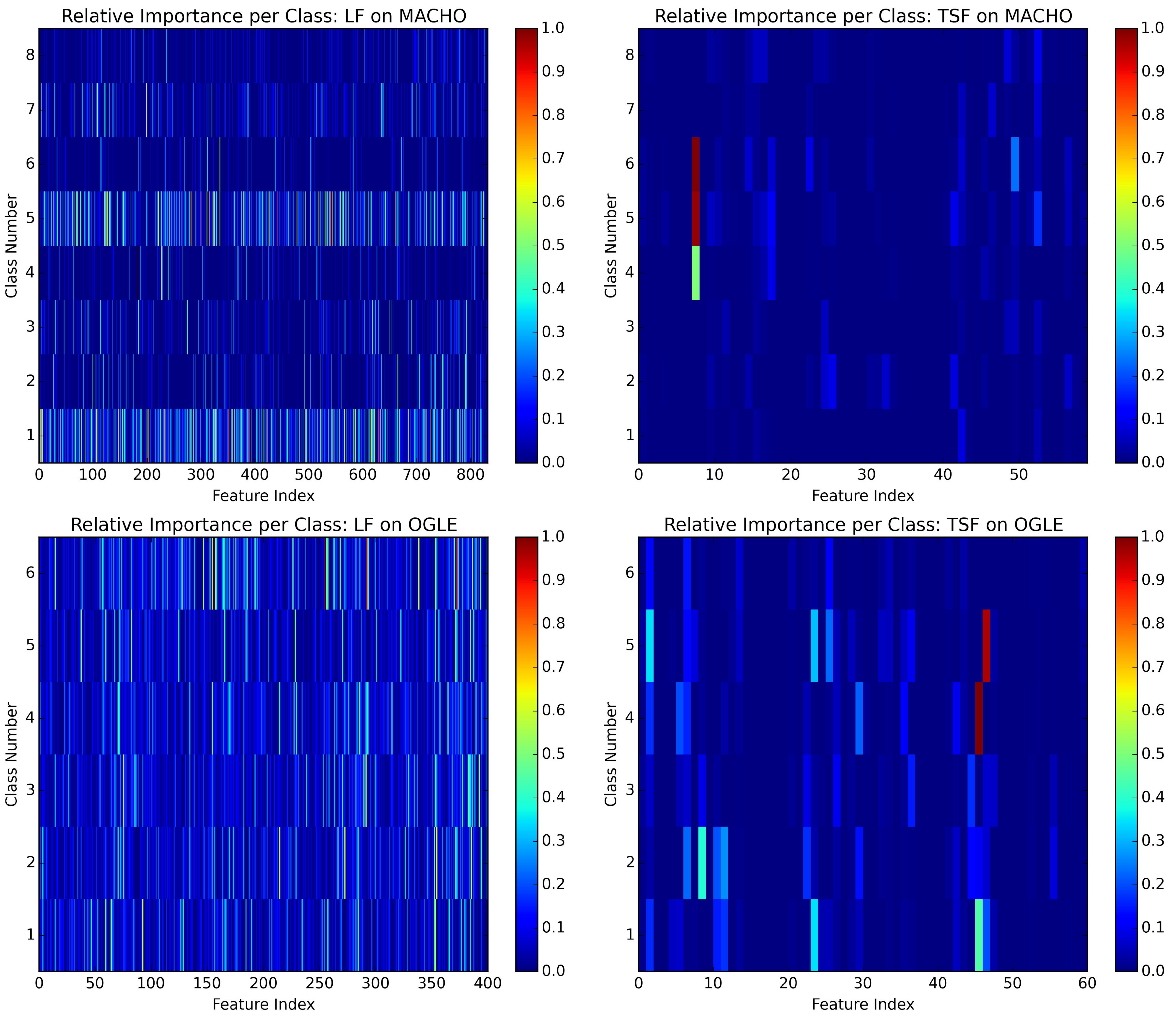}
\caption{Relative importance per class. The two heatmaps show the relative importance of each feature of both training sets constructed with the MACHO and OGLE data. We can see in the figure that the contribution of each learned feature to classification is complex while the contribution of designed features is in many cases minimal and classification with these features is largely based on the contribution of a few of the best features.}
\label{fig:heatmap}
\end{figure*}

Another way of analyzing the contribution of learned features to classification is looking at each of the features relative importance in each of the hyperplanes that are learned in each class during the SVM training (a multi-class SVM learns one hyperplane to separate each class from all of the rest). We can see in Figure \ref{fig:heatmap} that the contribution of learned features to classification is complex and that their contribution is different for each class: most of the features have very different relative importance values for each class. Time Series features, on the other hand, rely heavily on a few features for classification: a clear example of this is the red color of one feature for classes 5 and 6 in the top right heatmap, while the other features look mostly dark blue (the lowest relative importance value).

\section{Computational Run Time Analysis}
\label{sec:computational_cost}

To address the scalability requirements of future astronomical surveys, it is important that analysis algorithms run within manageable time frames and scale well with an increase in the volume of input data. Table \ref{table:run_time} shows approximate the run times for each of our method steps described in Section \ref{sec:method}. The two main algorithms on which we rely are Affinity Propagation and the Time Warp Edit Distance. The algorithmic complexity for Affinity Propagation is $O(N^2)$ where $N$ is the number of points (lightcurve subsequences) being clustered and the complexity of TWED is $O(pq)$ where $p$ and $q$ are the number of samples in each of the time series under comparison.

\begin{table}[H]
\caption{\label{table:run_time}Computational run time details.}
\centering
\begin{tabular}{| c | c |}
\hline
Step & Run Time \\
\hline
Sampling (Sec. \ref{sec:sampling}) & 5 minutes \\
Clustering (Sec. \ref{sec:clustering}) & 10 minutes \\
Encoding (Sec. \ref{sec:encoding}) & 1.5 - 3 hours \\
\hline
Total & 1.75 - 3.25 hours \\
\hline
\end{tabular}
\end{table}

Our method is also significantly faster in transforming a lightcurve from its time series to its encoded vector representation. If we compare against calculating the time series features we used for comparison in our experiments, we find that the speed gain is almost an order of magnitude. One might argue that this is not a fair comparison since our method depends on the execution of previous steps, namely sampling and clustering, but that overhead is a constant cost that doesn't increase with the number of lightcurves to be transformed.

\begin{table}[H]
\caption{\label{table:encoding_time}Average encoding time.}
\centering
\begin{tabular}{| c | c | c |}
\hline
Training Set & LF & TSF \\
\hline
MACHO & 1.95 s & 12.94 s \\
OGLE-III & 0.86 s & 7.70 s \\
\hline
\end{tabular}
\end{table}

\section{Conclusions}
\label{sec:conclusions}
In this work, we have introduced a new way of modeling and representing lightcurve data as input for automatic classifiers. The method does not assume previous knowledge about the lightcurves or use any expert knowledge, unlike previous traditional methods that use a set of timeseries features specially designed by astronomers for the task of classification. The previous fact together with the possibility of leveraging the vast amount of information available in unlabeled data constitutes a big step towards a more automatic, flexible and powerful classification pipeline. Our method works by extracting a large number of lightcurve subsequences from a given set of photometric data, which are then clustered to find common local patterns in the time series. Representatives of these common patterns, called exemplars, are then used to transform lightcurves of a labeled set into a new representation that can then be used to train an automatic classifier.

Our results show that this representation is as suitable for classification purposes than the traditional time series feature-based representation. Classifiers trained with our features perform as well as ones trained with expert designed features, while the computational cost of our method is significantly lower. With our method we were able to find 113,873 new variable star candidates. Our hope is that the research community will hold feature learning methods as a valid alternative to lightcurve representation in future work since we have shown them to be a strong competitor to the expert designed time series features. Our implementation code is readily available for others to download and build upon: users should try and adjust the parameters mentioned in Table \ref{table:parameters} to suit their particular application.

\section*{Acknowledgments}
This work is supported by Vicerrector\'ia de Investigaci\'on (VRI) from Pontificia Universidad Cat\'olica de Chile, Institute of Applied Computer Science at Harvard University, and the Chilean Ministry for the Economy, Development, and Tourism's Programa Iniciativa Cient\'{i}fica Milenio through grant P07-021-F, awarded to The Milky Way Millennium Nucleus. This research has made use of the SIMBAD database, operated at CDS, Strasbourg, France.

%\begin{thebibliography}{
%\bibliography{../../Bibtex/library,../../Documents/BibTex/library}
\bibliography{references_cm}

\begin{thebibliography}{63}
\expandafter\ifx\csname natexlab\endcsname\relax\def\natexlab#1{#1}\fi

\bibitem[{Alcock {et~al.}(1997)Alcock, Allsman, Alves, Axelrod, Bennett, Cook,
  Freeman, Griest, Guern, Lehner, {et~al.}}]{alcock1997macho}
Alcock, C., Allsman, R., Alves, D., Axelrod, T., Bennett, D., Cook, K.,
  Freeman, K., Griest, K., Guern, J., Lehner, M., {et~al.} 1997, The
  Astrophysical Journal, 479, 119

\bibitem[{Babu \& Mahabal(2015)}]{babu2015skysurveys}
Babu, G.~J., \& Mahabal, A. 2015, International Statistical Review

\bibitem[{Bell \& Sejnowski(1997)}]{bell1997independent}
Bell, A.~J., \& Sejnowski, T.~J. 1997, Vision research, 37, 3327

\bibitem[{Bengio(2009)}]{bengio2009learning}
Bengio, Y. 2009, Foundations and Trends in Machine Learning, 2, 1

\bibitem[{Berndt \& Clifford(1994)}]{berndt1994using}
Berndt, D.~J., \& Clifford, J. 1994, in KDD workshop, Vol.~10, Seattle, WA,
  359--370

\bibitem[{Bloom \& Richards(2011)}]{bloom2011data}
Bloom, J., \& Richards, J. 2011, Advances in Machine Learning and Data Mining
  for Astronomy

\bibitem[{Bloom {et~al.}(2012)Bloom, Richards, Nugent, Quimby, Kasliwal, Starr,
  Poznanski, Ofek, Cenko, Butler, {et~al.}}]{bloom2012automating}
Bloom, J., Richards, J., Nugent, P., Quimby, R., Kasliwal, M., Starr, D.,
  Poznanski, D., Ofek, E., Cenko, S., Butler, N., {et~al.} 2012, Publications
  of the Astronomical Society of the Pacific, 124, 1175

\bibitem[{Boser {et~al.}(1992)Boser, Guyon, \& Vapnik}]{boser1992training}
Boser, B.~E., Guyon, I.~M., \& Vapnik, V.~N. 1992, in Proceedings of the fifth
  annual workshop on Computational learning theory, ACM, 144--152

\bibitem[{Boureau {et~al.}(2010)Boureau, Ponce, \&
  LeCun}]{boureau2010theoretical}
Boureau, Y.-L., Ponce, J., \& LeCun, Y. 2010, in Proceedings of the 27th
  International Conference on Machine Learning (ICML-10), 111--118

\bibitem[{Breiman(2001)}]{Breiman2001}
Breiman, L. 2001, Machine learning, 5

\bibitem[{Chen {et~al.}(2005)Chen, {\"O}zsu, \& Oria}]{chen2005robust}
Chen, L., {\"O}zsu, M.~T., \& Oria, V. 2005, in Proceedings of the 2005 ACM
  SIGMOD international conference on Management of data, ACM, 491--502

\bibitem[{Coates \& Ng(2012)}]{coates2012learning}
Coates, A., \& Ng, A.~Y. 2012, in Neural Networks: Tricks of the Trade
  (Springer), 561--580

\bibitem[{Cortes \& Vapnik(1995)}]{Cortes:1995}
Cortes, C., \& Vapnik, V. 1995, Machine Learning, 20, 273

\bibitem[{Dahl {et~al.}(2010)Dahl, Mohamed, Hinton, {et~al.}}]{dahl2010phone}
Dahl, G., Mohamed, A.-r., Hinton, G.~E., {et~al.} 2010, in Advances in neural
  information processing systems, 469--477

\bibitem[{Dahl {et~al.}(2012)Dahl, Yu, Deng, \& Acero}]{dahl2012context}
Dahl, G.~E., Yu, D., Deng, L., \& Acero, A. 2012, Audio, Speech, and Language
  Processing, IEEE Transactions on, 20, 30

\bibitem[{Debosscher {et~al.}(2007)Debosscher, Sarro, Aerts, Cuypers,
  Vandenbussche, Garrido, \& Solano}]{Debosscher:2007}
Debosscher, J., Sarro, L., Aerts, C., Cuypers, J., Vandenbussche, B., Garrido,
  R., \& Solano, E. 2007, Astronomy and Astrophysics, 475, 1159

\bibitem[{Frey \& Dueck(2007)}]{frey2007clustering}
Frey, B.~J., \& Dueck, D. 2007, science, 315, 972

\bibitem[{Graves {et~al.}(2013)Graves, Mohamed, \& Hinton}]{graves2013speech}
Graves, A., Mohamed, A.-r., \& Hinton, G. 2013, in Acoustics, Speech and Signal
  Processing (ICASSP), 2013 IEEE International Conference on, IEEE, 6645--6649

\bibitem[{Grosse {et~al.}(2007)Grosse, Raina, Kwong, \& Ng}]{Grosse07}
Grosse, R., Raina, R., Kwong, H., \& Ng, A. 2007, in Proceedings of the
  Twenty-Third Conference Annual Conference on Uncertainty in Artificial
  Intelligence (UAI-07) (Corvallis, Oregon: AUAI Press), 149--158

\bibitem[{Guyon {et~al.}(2002)Guyon, Weston, Barnhill, \&
  Vapnik}]{guyon2002gene}
Guyon, I., Weston, J., Barnhill, S., \& Vapnik, V. 2002, Machine learning, 46,
  389

\bibitem[{Hanif \& Protopapas(2015)}]{hanif2015recursive}
Hanif, A., \& Protopapas, P. 2015, Monthly Notices of the Royal Astronomical
  Society, 448, 390

\bibitem[{Hinton {et~al.}(2006)Hinton, Osindero, \& Teh}]{hinton2006fast}
Hinton, G.~E., Osindero, S., \& Teh, Y.-W. 2006, Neural computation, 18, 1527

\bibitem[{Hinton \& Salakhutdinov(2006)}]{hinton2006reducing}
Hinton, G.~E., \& Salakhutdinov, R.~R. 2006, Science, 313, 504

\bibitem[{Huijse {et~al.}(2012)Huijse, Est{\'e}vez, Protopapas, Zegers,
  Principe, {et~al.}}]{huijse2012information}
Huijse, P., Est{\'e}vez, P., Protopapas, P., Zegers, P., Principe, J.~C.,
  {et~al.} 2012, Signal Processing, IEEE Transactions on, 60, 5135

\bibitem[{H{\"u}sken \& Stagge(2003)}]{husken2003recurrent}
H{\"u}sken, M., \& Stagge, P. 2003, Neurocomputing, 50, 223

\bibitem[{Jaitly \& Hinton(2011)}]{jaitly2011learning}
Jaitly, N., \& Hinton, G. 2011, in Acoustics, Speech and Signal Processing
  (ICASSP), 2011 IEEE International Conference on, IEEE, 5884--5887

\bibitem[{Kaiser {et~al.}(2002)Kaiser, Aussel, Burke, Boesgaard, Chambers,
  Chun, Heasley, Hodapp, Hunt, Jedicke, Jewitt, Kudritzki, Luppino, Maberry,
  Magnier, Monet, Onaka, Pickles, Rhoads, Simon, Szalay, Szapudi, Tholen,
  Tonry, Waterson, \& Wick}]{Kaiser2002SPIE}
Kaiser, N., Aussel, H., Burke, B.~E., Boesgaard, H., Chambers, K., Chun, M.~R.,
  Heasley, J.~N., Hodapp, K.-W., Hunt, B., Jedicke, R., Jewitt, D., Kudritzki,
  R., Luppino, G.~A., Maberry, M., Magnier, E., Monet, D.~G., Onaka, P.~M.,
  Pickles, A.~J., Rhoads, P.~H.~H., Simon, T., Szalay, A., Szapudi, I., Tholen,
  D.~J., Tonry, J.~L., Waterson, M., \& Wick, J. 2002, in Society of
  Photo-Optical Instrumentation Engineers (SPIE) Conference Series, Vol. 4836,
  Society of Photo-Optical Instrumentation Engineers (SPIE) Conference Series,
  ed. {J.\~{}A.\~{}Tyson \& S.\~{}Wolff}, 154--164

\bibitem[{Keller {et~al.}(2007)Keller, Schmidt, Bessell, Conroy, Francis,
  Granlund, Kowald, Oates, Martin-Jones, Preston, Tisserand, Vaccarella, \&
  Waterson}]{Keller:2007}
Keller, S.~C., Schmidt, B.~P., Bessell, M.~S., Conroy, P.~G., Francis, P.,
  Granlund, A., Kowald, E., Oates, A.~P., Martin-Jones, T., Preston, T.,
  Tisserand, P., Vaccarella, A., \& Waterson, M.~F. 2007, Publications of the
  Astronomical Society of Australia, 24

\bibitem[{Keogh \& Lin(2005)}]{keogh2005clustering}
Keogh, E., \& Lin, J. 2005, Knowledge and information systems, 8, 154

\bibitem[{Kim {et~al.}(2009)Kim, Protopapas, Alcock, Byun, \&
  Bianco}]{kim2009detrending}
Kim, D.-W., Protopapas, P., Alcock, C., Byun, Y.-I., \& Bianco, F.~B. 2009,
  Monthly Notices of the Royal Astronomical Society, 397, 558

\bibitem[{Kim {et~al.}(2014)Kim, Protopapas, Bailer-Jones, Byun, Chang,
  Marquette, \& Shin}]{kim2014epoch}
Kim, D.-W., Protopapas, P., Bailer-Jones, C.~A., Byun, Y.-I., Chang, S.-W.,
  Marquette, J.-B., \& Shin, M.-S. 2014, Astronomy \& Astrophysics, 566, A43

\bibitem[{Kim {et~al.}(2011)Kim, Protopapas, Byun, Alcock, Khardon, \&
  Trichas}]{Kim:2011}
Kim, D.-W., Protopapas, P., Byun, Y.-I., Alcock, C., Khardon, R., \& Trichas,
  M. 2011, The Astrophysical Journal, 735

\bibitem[{Kim {et~al.}(2012)Kim, Protopapas, Trichas, Rowan-Robinson, Khardon,
  Alcock, \& Byun}]{Kim:2012}
Kim, D.-W., Protopapas, P., Trichas, M., Rowan-Robinson, M., Khardon, R.,
  Alcock, C., \& Byun, Y.-I. 2012, The Astrophysical Journal, 747

\bibitem[{Krizhevsky {et~al.}(2010)Krizhevsky, Hinton,
  {et~al.}}]{krizhevsky2010factored}
Krizhevsky, A., Hinton, G.~E., {et~al.} 2010, in International Conference on
  Artificial Intelligence and Statistics, 621--628

\bibitem[{L{\"a}ngkvist \& Loutfi(2012)}]{langkvist2012not}
L{\"a}ngkvist, M., \& Loutfi, A. 2012, in NIPS workshop on Deep Learning and
  Unsupervised Feature Learning

\bibitem[{Larochelle \& Bengio(2008)}]{larochelle2008classification}
Larochelle, H., \& Bengio, Y. 2008, in Proceedings of the 25th international
  conference on Machine learning, ACM, 536--543

\bibitem[{Lee {et~al.}(2006)Lee, Battle, Raina, \& Ng}]{lee2006efficient}
Lee, H., Battle, A., Raina, R., \& Ng, A.~Y. 2006, in Advances in neural
  information processing systems, 801--808

\bibitem[{Levenshtein(1966)}]{levenshtein1966binary}
Levenshtein, V. 1966, Soviet Physics Doklady, 10, 707

\bibitem[{Marteau(2009)}]{marteau2009time}
Marteau, P.-F. 2009, Pattern Analysis and Machine Intelligence, IEEE
  Transactions on, 31, 306

\bibitem[{Masci {et~al.}(2014)Masci, Hoffman, Grillmair, \&
  Cutri}]{masci2014automated}
Masci, F.~J., Hoffman, D.~I., Grillmair, C.~J., \& Cutri, R.~M. 2014, The
  Astronomical Journal, 148, 21

\bibitem[{Matter(2007)}]{Matter:2007}
Matter, D. 2007, Science, 1

\bibitem[{McKinney(2010)}]{mckinney2010data}
McKinney, W. 2010, in Proceedings of the 9th, Vol. 445, 51--56

\bibitem[{Mohamed {et~al.}(2012)Mohamed, Dahl, \& Hinton}]{mohamed2012acoustic}
Mohamed, A.-r., Dahl, G.~E., \& Hinton, G. 2012, Audio, Speech, and Language
  Processing, IEEE Transactions on, 20, 14

\bibitem[{Nam(2012)}]{nam2012learning}
Nam, J. 2012, PhD thesis, Stanford University

\bibitem[{Neff {et~al.}(2015)Neff, Wells, Geltz, \& Brown}]{neff2015automated}
Neff, J., Wells, M., Geltz, S., \& Brown, A. 2015, in Cambridge Workshop on
  Cool Stars, Stellar Systems, and the Sun, Vol.~18, 879--886

\bibitem[{Nun {et~al.}(2014)Nun, Pichara, Protopapas, \&
  Kim}]{nun2014supervised}
Nun, I., Pichara, K., Protopapas, P., \& Kim, D.-W. 2014, The Astrophysical
  Journal, 793, 23

\bibitem[{Nun {et~al.}(2015)Nun, Protopapas, Sim, Zhu, Dave, Castro, \&
  Pichara}]{nun2015fats}
Nun, I., Protopapas, P., Sim, B., Zhu, M., Dave, R., Castro, N., \& Pichara, K.
  2015, arXiv preprint arXiv:1506.00010

\bibitem[{Olshausen {et~al.}(1996)}]{olshausen1996emergence}
Olshausen, B.~A., {et~al.} 1996, Nature, 381, 607

\bibitem[{Paczynski(1986)}]{paczynski1986gravitational}
Paczynski, B. 1986, The Astrophysical Journal, 304, 1

\bibitem[{Pedregosa {et~al.}(2011)Pedregosa, Varoquaux, Gramfort, Michel,
  Thirion, Grisel, Blondel, Prettenhofer, Weiss, Dubourg, Vanderplas, Passos,
  Cournapeau, Brucher, Perrot, \& Duchesnay}]{scikit-learn}
Pedregosa, F., Varoquaux, G., Gramfort, A., Michel, V., Thirion, B., Grisel,
  O., Blondel, M., Prettenhofer, P., Weiss, R., Dubourg, V., Vanderplas, J.,
  Passos, A., Cournapeau, D., Brucher, M., Perrot, M., \& Duchesnay, E. 2011,
  Journal of Machine Learning Research, 12, 2825

\bibitem[{Pichara \& Protopapas(2013)}]{Pichara_Miss:2013}
Pichara, K., \& Protopapas, P. 2013, The Astrophysical Journal, 777, 83

\bibitem[{Pichara {et~al.}(2012{\natexlab{a}})Pichara, Protopapas, Kim,
  Marquette, \& Tisserand}]{Pichara_QSO:2012}
Pichara, K., Protopapas, P., Kim, D., Marquette, J., \& Tisserand, P.
  2012{\natexlab{a}}, Monthly Notices of the Royal Academy Society, 18, 1

\bibitem[{Pichara {et~al.}(2012{\natexlab{b}})Pichara, Protopapas, Kim,
  Marquette, \& Tisserand}]{pichara2012improved}
Pichara, K., Protopapas, P., Kim, D.-W., Marquette, J.-B., \& Tisserand, P.
  2012{\natexlab{b}}, Monthly Notices of the Royal Astronomical Society, 427,
  1284

\bibitem[{Poultney {et~al.}(2006)Poultney, Chopra, Cun,
  {et~al.}}]{poultney2006efficient}
Poultney, C., Chopra, S., Cun, Y.~L., {et~al.} 2006, in Advances in neural
  information processing systems, 1137--1144

\bibitem[{Ranzato \& Hinton(2010)}]{ranzato2010modeling}
Ranzato, M., \& Hinton, G.~E. 2010, in Computer Vision and Pattern Recognition
  (CVPR), 2010 IEEE Conference on, IEEE, 2551--2558

\bibitem[{Rebbapragada {et~al.}(2009)Rebbapragada, Protopapas, Brodley, \&
  Alcock}]{rebbapragada2009finding}
Rebbapragada, U., Protopapas, P., Brodley, C.~E., \& Alcock, C. 2009, Machine
  learning, 74, 281

\bibitem[{Richards {et~al.}(2011)Richards, Starr, Butler, Bloom, Brewer,
  Crellin-Quick, Higgins, Kennedy, \& Rischard}]{Richards:2011}
Richards, J.~W., Starr, D.~L., Butler, N.~R., Bloom, J.~S., Brewer, J.~M.,
  Crellin-Quick, A., Higgins, J., Kennedy, R., \& Rischard, M. 2011, The
  Astrophysical Journal, 733

\bibitem[{Serr{\`a} \& Arcos(2014)}]{serra2014empirical}
Serr{\`a}, J., \& Arcos, J.~L. 2014, Knowledge-Based Systems, 67, 305

\bibitem[{Udalski {et~al.}(2008)Udalski, Szymanski, Soszynski, \&
  Poleski}]{Udalski2008AcA}
Udalski, A., Szymanski, M.~K., Soszynski, I., \& Poleski, R. 2008, Acta
  Astronomica, 58, 69

\bibitem[{van~der Walt {et~al.}(2011)van~der Walt, Colbert, \&
  Varoquaux}]{numpy-scipy}
van~der Walt, S., Colbert, S.~C., \& Varoquaux, G. 2011, CoRR, abs/1102.1523

\bibitem[{Wachman {et~al.}(2009)Wachman, Khardon, Protopapas, \&
  Alcock}]{Wachman:2009}
Wachman, G., Khardon, R., Protopapas, P., \& Alcock, C. 2009, in Lecture Notes
  in Computer Science, Vol. 5782, Machine Learning and Knowledge Discovery in
  Databases, ed. W.~Buntine, M.~Grobelnik, D.~Mladenic, \& J.~Shawe-Taylor
  (Springer Berlin / Heidelberg), 489--505

\bibitem[{Wang {et~al.}(2010)Wang, Khardon, \& Protopapas}]{Wang:2010}
Wang, Y., Khardon, R., \& Protopapas, P. 2010, in Lecture Notes in Computer
  Science, Vol. 6323, Machine Learning and Knowledge Discovery in Databases
  (Springer Berlin / Heidelberg), 418--434

\bibitem[{Wenger {et~al.}(2000)Wenger, Ochsenbein, Egret, Dubois, Bonnarel,
  Borde, Genova, Jasniewicz, Lalo{\"e}, Lesteven, {et~al.}}]{wenger2000simbad}
Wenger, M., Ochsenbein, F., Egret, D., Dubois, P., Bonnarel, F., Borde, S.,
  Genova, F., Jasniewicz, G., Lalo{\"e}, S., Lesteven, S., {et~al.} 2000,
  Astronomy and Astrophysics Supplement Series, 143, 9

\end{thebibliography}
%\end{thebibliography}

%\appendix
%\section{Arc Reversal}
%\label{app:ArcRev}

\end{document}